\documentclass[%
journal=cmatex,
doi=true,
manuscript=article,layout=onecolumn
]{achemso}

\usepackage{graphicx}%
\usepackage{dcolumn}%
\usepackage{bm}%
\usepackage{verbatim}
\usepackage{isomath}
\usepackage{bbold}
\usepackage{cancel}
\usepackage{mathrsfs}
\usepackage{bbm}
\usepackage{amssymb,graphicx}
\usepackage{physics}
\usepackage{float}
\usepackage{parskip}
\usepackage[fleqn]{mathtools}
\usepackage[names]{xcolor}
\allowdisplaybreaks

\usepackage{xr}
\makeatletter
\newcommand*{\addFileDependency}[1]{
  \typeout{(#1)}
  \@addtofilelist{#1}
  \IfFileExists{#1}{}{\typeout{No file #1.}}
}
\makeatother

\newcommand*{\myexternaldocument}[1]{
    \externaldocument{#1}
    \addFileDependency{#1.tex}
    \addFileDependency{#1.aux}
}

\myexternaldocument{si}

\def\vector#1{\mathbf{#1}}
\def\tensor#1{\tensorsym{#1}}

\def\der#1{\textrm{d} #1~}
\def\Det#1{\textrm{det} \left(#1\right)}
\def\avg#1{\left< #1 \right>}

\title{Modeling the structural and thermal properties of loaded metal-organic frameworks.\\An interplay of quantum and anharmonic fluctuations}
\author{Venkat Kapil}
\altaffiliation{These authors contributed equally to this work}
\affiliation{Laboratory of Computational Science and Modelling, Institute of Materials, Ecole Polytechnique F\'ed\'erale de Lausanne, Lausanne, Switzerland
}%

\author{Jelle Wieme}
\altaffiliation{These authors contributed equally to this work}
\affiliation{Center for Molecular Modeling, Ghent University, Tech Lane Ghent Science Park Campus A, Technologiepark 903, 9052 Zwijnaarde, Belgium}%
\author{Steven Vandenbrande}
\author{Aran Lamaire}
\author{Veronique Van Speybroeck}
\affiliation{Center for Molecular Modeling, Ghent University, Tech Lane Ghent Science Park Campus A, Technologiepark 903, 9052 Zwijnaarde, Belgium}%

\author{Michele Ceriotti}
\email{michele.ceriotti@epfl.ch}
\affiliation{Laboratory of Computational Science and Modelling, Institute of Materials, Ecole Polytechnique F\'ed\'erale de Lausanne, Lausanne, Switzerland
}%
\date{\today}%

\begin{document}

\begin{abstract}

Metal-organic frameworks show both fundamental interest and great promise for applications in adsorption-based technologies, such as the separation and storage of gases. 
The flexibility and complexity of the molecular scaffold poses a considerable challenge to atomistic modeling, especially when also considering the presence of guest molecules. 
We investigate the role played by quantum and anharmonic fluctuations in the archetypical case of MOF-5, comparing the material at various levels of methane loading.
Accurate path integral simulations of such effects are made affordable by the introduction of an accelerated simulation scheme and the use of an optimized force field based on first-principles reference calculations.
We find that the level of statistical treatment that is required for predictive modeling depends significantly on the property of interest. The thermal properties of the lattice are generally well described by a quantum harmonic treatment, with the adsorbate behaving in a classical but strongly anharmonic manner. 
The heat capacity of the loaded framework  -- which plays an important role in the characterization of the framework and in determining its stability to thermal fluctuations during adsorption/desorption cycles -- requires, however, a full quantum and anharmonic treatment, either by path integral methods or by a simple but approximate scheme. 
We also present molecular-level insight into the nanoscopic interactions contributing to the material's properties and suggest design principles to optimize them. 
\end{abstract}

\maketitle

\section{\label{sec:introduction} INTRODUCTION}

Tailor-made porous materials \cite{slater2015} like metal-organic frameworks (MOFs) \cite{furukawa2013} are at the core of emerging technologies due to their exceptional physical and chemical properties such as a tunable ultrahigh porosity and an associated enormous gas storage capacity.
Therefore, they have been proposed for applications such as adsorbed natural gas (ANG) storage in vehicles \cite{mason2014,he2017},  adsorption-driven heat pumps \cite{delange2015,wang2018}, and carbon capture and sequestration (CCS) \cite{sumida2012,trickett2017}. 
While a lot of work still needs to be done to optimize the crucial adsorption and storage properties of these porous materials, \cite{simon2015} studies on other critical requirements such as heat management are gaining interest. \cite{mason2016}
For instance, the heat capacity, \textit{i.e.}, the amount of energy required to increase the material's temperature, is a fundamental thermodynamic property of interest in these applications which involve large thermal fluctuations as adsorption and desorption processes imply the release or consumption of energy. Moreover, the heat capacity of the MOF affects the energy penalty to regenerate the adsorbent in, for example, CCS. \cite{huck2014} 
To date, however, information on the heat capacity is lacking for most MOFs \cite{mu2011,kloutse2015} and the influence of adsorbed guest molecules on the heat capacity has not yet been investigated, in contrast to other thermal properties such as the thermal expansion behavior and the thermal conductivity. \cite{babaei2016, balestra2016,auckett2018} 

Within this context, an efficient and accurate simulation protocol to tackle the structural and thermal properties of MOFs including all the relevant physical effects could facilitate a better understanding of the structure-property relations and suggest design principles for materials with improved properties. Due to the importance of finite-temperature effects, anharmonicity, and nuclear quantum effects (NQEs), the modeling of the thermophysics of MOFs is generally not a trivial exercise. The first two effects have already been the subject of many investigations and were included in our protocol to characterize the thermodynamics of MOFs. \cite{rogge2015,vanduyfhuys2018b} Furthermore, very recently, some of the present authors highlighted the necessity of an accurate theoretical framework for the design of thermoresponsive MOFs. \cite{wieme2018} However, the impact of NQEs has so far received far less attention within the MOF community, despite the many light atoms contained in the crystal structure and present inside the pores. \cite{paesani2012,borges2017}

In this regard, path integral molecular dynamics (PIMD) \cite{parr-rahm84jcp} provides an ideal reference framework for the evaluation of thermodynamic averages, as it seamlessly captures both NQEs and the anharmonic motion of nuclei. The statistics of distinguishable quantum particles can be obtained through the equivalence between the thermodynamics of a quantum system and a classical ring polymer containing $P$ replicas of the system. \cite{chan-woly81jcp} In the limit of large $P$ values, NQEs can then systematically be accounted for. The major downside of this technique is the associated high computational cost, \textit{i.e.}, $P$ times the cost of classical molecular dynamics (MD). However, several methodological advances \cite{jang-voth01jcp, ceri12prl,  kapi+16jcp2, mark-mano08jcp, kapi+16jcp, polt-tkat16cs} that enable a reduction of the computational cost  have made it a mainstream technique for material modeling. \cite{mark-ceri18nrc}

\begin{figure*}[t]
	\begin{center}
		\includegraphics[width=\linewidth]{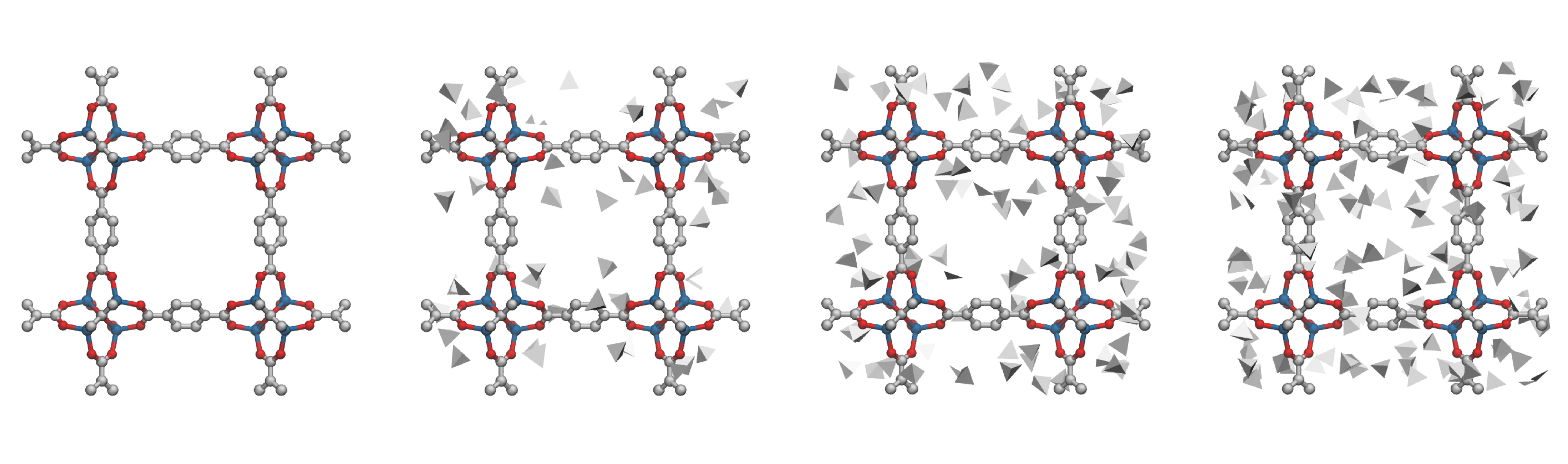}
	\end{center}
	\caption{\label{fig:mof_pretty_structures} The structure of the MOF-5 with a gas loading $x$ of 0, 50, 100, and 150 molecules of methane (left to right) in the conventional unit cell. (8(Zn$_4$O(CO$_2$)$_6$) $\cdot$ $x$ CH$_4$) The oxygen, carbon, and zinc atoms are shown in red, silver, and blue respectively. For the sake of aesthetics the hydrogen atoms are not included. The methane molecules are represented by silver tetrahedra. }  
\end{figure*}

An additional difficulty arises from the fact that most experiments and practical applications are performed in isothermal-isobaric conditions, while the vast majority of atomistic simulations are performed with a fixed unit cell, corresponding to isochoric conditions. As most solid materials have a very small compressibility, the difference between the two ensembles is often negligible. For MOFs on the contrary -- particularly when loaded with a gas -- the behavior in isobaric and isochoric conditions can be very different. Some of us emphasized the importance of taking into account the variations of the cell shape to simulate properties of flexible MOFs. \cite{rogge2015,Rogge2018} 
While algorithms for performing path integral simulations at constant pressure conditions exist~\cite{mart+99jcp, ipi1}, an accurate evaluation of the thermophysical properties requires a very large number of replicas for convergence. 
In this article, we introduce a method, based on a recently developed implementation of high order path integral factorizations~\cite{kapi+16jcp2}, to greatly accelerate the convergence of these simulations.

This method, in combination with a first-principles-based force field~\cite{vanduyfhuys2015,vanduyfhuys2018,vanduyfhuys2018b}, makes it possible to characterize the structural and thermophysical properties of complex molecular systems such as guest-loaded MOFs. We investigate the archetypical case of the well-known MOF-5 \cite{li1999,eddaoudi2002} in the presence and absence of methane in its pores (see Figure \ref{fig:mof_pretty_structures}). Evaluating and understanding the impact of methane adsorption on the properties of MOFs is especially important as they have been proposed as potential adsorbents for natural gas storage applications. \cite{mason2014,simon2015,he2017}
We demonstrate the crucial role of a complete statistical-mechanical description of the quantum and anharmonic fluctuations in MOFs for a correct description of structural properties and the heat capacity of guest-loaded MOFs. By meticulously disentangling anharmonic and nuclear quantum effects for both the lattice and the guest particles, we are able to propose an efficient empirical calculation scheme which may be used to screen MOFs with beneficial thermal properties on a larger scale. 

\begingroup
\section{\label{sec:methodology} METHODOLOGY}

\subsection{\label{sec:mater} Materials}
The materials that are considered in this theoretical work are pristine and methane-loaded MOF-5 scaffolds.~\cite{li1999} This framework consists of Zn$_4$O(CO$_2$)$_6$ inorganic nodes connected through 1,4-benzenedicarboxylate (bdc) linkers. The unit cell is cubic and contains eight inorganic nodes, as shown in Figure \ref{fig:mof_pretty_structures}. We consider three different loadings $x$ of $50$, $100$, and $150$ methane molecules in the conventional unit cell (8(Zn$_4$O(CO$_2$)$_6$) $\cdot$ $x$ CH$_4$), which encompasses both the low- and high-adsorption regime. \cite{zhou2007,mason2014} At 100 bar, for example, approximately 120 methane molecules are present per conventional unit cell, as measured by Mason \textit{et al.} \cite{mason2014} (see SI Figure S2).

\subsection{\label{ss:ff} First-principles-derived force fields}

The molecular simulations are performed using newly developed force fields for MOF-5 and methane. They are derived with \texttt{QuickFF}~\cite{vanduyfhuys2015,vanduyfhuys2018}, a software package developed to derive force fields for MOFs in an easy yet accurate way based on information obtained from first-principles input data. Isolated cluster models were used to generate the required first-principles input data, which includes the geometry and the Hessian in equilibrium together with the atomic charges. Within the \texttt{QuickFF} protocol, the quantum mechanical potential energy surface (PES) is approximated by a sum of analytical functions of the nuclear coordinates that describe the covalent and noncovalent interactions. The covalent interactions, which mimic the chemical bonds between the atoms, are approximated by different terms as a function of the internal coordinates (bonds, bends, out-of-plane distances, and dihedrals). The noncovalent interactions are composed of electrostatic and van der Waals interactions. The guest-host interactions between MOF-5 and methane only include noncovalent terms. A detailed discussion of the force field energy expression and derivation is provided in the Supporting Information (see SI Section S4).

\subsection{Thermodynamic ensembles}
\label{ss:thermo_ens}
In order to study the classical and quantum isobaric heat capacity of the guest-loaded MOF scaffold, the classical and quantum  isothermal-isobaric thermodynamic ensembles -- as defined in Ref.\ \cite{rogge2015} -- are respectively used.  For a system with  $N$ particles, subjected to an external mechanical stress $\tensor{\sigma} =  \mathcal{P} \tensor{I} + \tensor{\sigma}_{a}$, with $\mathcal{P}$ the  hydrostatic pressure and $\tensor{\sigma}_{a}$ the deviatoric stress, the classical $(\square = \text{cl})$ and quantum $(\square = \text{qn})$ isothermal-isobaric ensembles at temperature $k_BT = \beta^{-1}$ and fixed normalized cell tensor $\tensor{h}_{0}$ are described by the partition functions
\begin{align}
\Delta_{\square}(N, \mathcal{P}(\tensor{h}_{0}), T) \propto \int \der{\mathcal{V}} e^{-\beta \mathcal{P} \mathcal{V}} Z_{\square}(N,\mathcal{V}(\tensor{h}_{0}), T),
\end{align}
where $Z_{\square}(N,\mathcal{V}(\tensor{h}_{0}), T)$ are the corresponding canonical partition functions at volume $\mathcal{V}$, normalized cell tensor $\tensor{h}_{0}$, and temperature $T$. Similarly, the flexible $N\mathcal{P}(\tensor{\sigma}_a=\mathbf{0})T$ partition functions, which allow the cell shape to change, are defined by the partition functions \cite{mart+94jcp}:
\begin{align}
    \begin{split}
        & \Delta_{\square}(N, \mathcal{P}(\tensor{\sigma}_{a} = \mathbf{0}), T) \propto \\
& \int \der{\mathcal{V}} e^{-\beta \mathcal{P} \mathcal{V}} \int \der{\tensor{h}_0} \delta(\Det{\tensor{h}_0} - 1) Z_{\square}(N,\mathcal{V}(\tensor{h}_{0}), T)
    \end{split}.
\end{align}
 The classical ensembles are sampled by classical molecular dynamics \cite{Rahman1964}, whereas the quantum ensembles are sampled using path integral molecular dynamics \cite{parr-rahm84jcp}. The latter maps the quantum partition function of a system to that of a classical ring polymer Hamiltonian made of $P$ replicas of the system. \cite{chan-woly81jcp} 
Formally, exact results can be obtained in the limit of $P \to \infty$. Unfortunately, the requisite value of $P$ to calculate structural properties rises rapidly with decreasing temperature \cite{uhl+16jcp} and for properties such as heat capacity \cite{shiga2005} the increase in $P$ is even greater. 
This makes the standard scheme prohibitively expensive. 

\subsection{\label{ss:acpimd} Accelerated simulation scheme}

To remedy this problem, we present an accelerated scheme based on a constant pressure integrator for the high order path integral method \cite{jang-voth01jcp}. The quantum statistics of distinguishable particles arises from the non-commutative nature of the potential and kinetic energy operators. In standard (second order) path integral schemes, an approximate factorization of the high-temperature Boltzmann operator is introduced, that leads to an error that decreases as $\mathcal{O}(1/P^2)$.
High order techniques use an alternative splitting of the Boltzmann operator \cite{taka-imad84jpsj,chin97pla}, leading to an alternative ring polymer Hamiltonian with a faster, $\mathcal{O}(1/P^4)$ convergence to the exact quantum limit. This makes it possible to reduce the number of replicas and hence the computational cost. \cite{jang-voth01jcp} 
While many high order schemes exist \cite{taka-imad84jpsj,polt-tkat16cs}, here we focus on the specific case \cite{alpha_parameter} of a fourth-order Suzuki-Chin (SC) splitting \cite{suzu95pla,chin97pla}, which yields a so-called SC Hamiltonian $H_P^{\text{sc}}\left(\vector{p},\vector{q}\right)$. 
Considering for simplicity the case of a single particle in an external potential,
the SC Hamiltonian takes the form:
\begin{align}
H_P^{\text{sc}}\left(\vector{p},\vector{q}\right) = H_P^{0}\left(\vector{p},\vector{q}\right) + V_P^{\text{sc}}\left(\vector{q}\right),
\end{align}
where 
\begin{align*}
H_P^{0}\left(\vector{p},\vector{q}\right) = \sum_{j=1}^{P} \frac{\left[\vector{p}^{(j)}\right]^2}{2 m} +  \sum_{j=1}^{P} \frac{1}{2} m \omega_P^{2} \left[\vector{q}^{(j)} -\vector{q}^{(j+1)}\right]^2 
\end{align*}
is the ring polymer Hamiltonian of a free particle, subjected to cyclic boundary conditions ($j+P = j$), and \begin{align}
V_P^{\text{sc}}\left(\vector{q}\right) & = \sum_{j=1}^{P/2}  \left[\frac{2}{3} V\left(\vector{q}^{(2j-1)}\right) + \frac{4}{3} V\left(\vector{q}^{(2j)}\right)+ \frac{1}{9} \tilde{V}\left(\vector{q}^{(2j)}\right) \right].
\end{align}
Note that the odd and even replicas feel the physical potential $V\left(\vector{q}^{(j)}\right)$ scaled by factors of $2/3$ and $4/3$ respectively and that the high order term  $\tilde{V}\left(\vector{q}^{(j)}\right) = \omega_P^{-2} m^{-1}|\vector{f}^{(j)}|^{2}$, that depends on the modulus of the force $\vector{f}^{(j)}\equiv - {\partial V\left(\vector{q}^{(j)}\right)} / {\partial \vector{q}^{(j)}}$, only acts on the even replicas. 

\begin{figure}[t]
\centering\includegraphics[width= 0.50\linewidth]{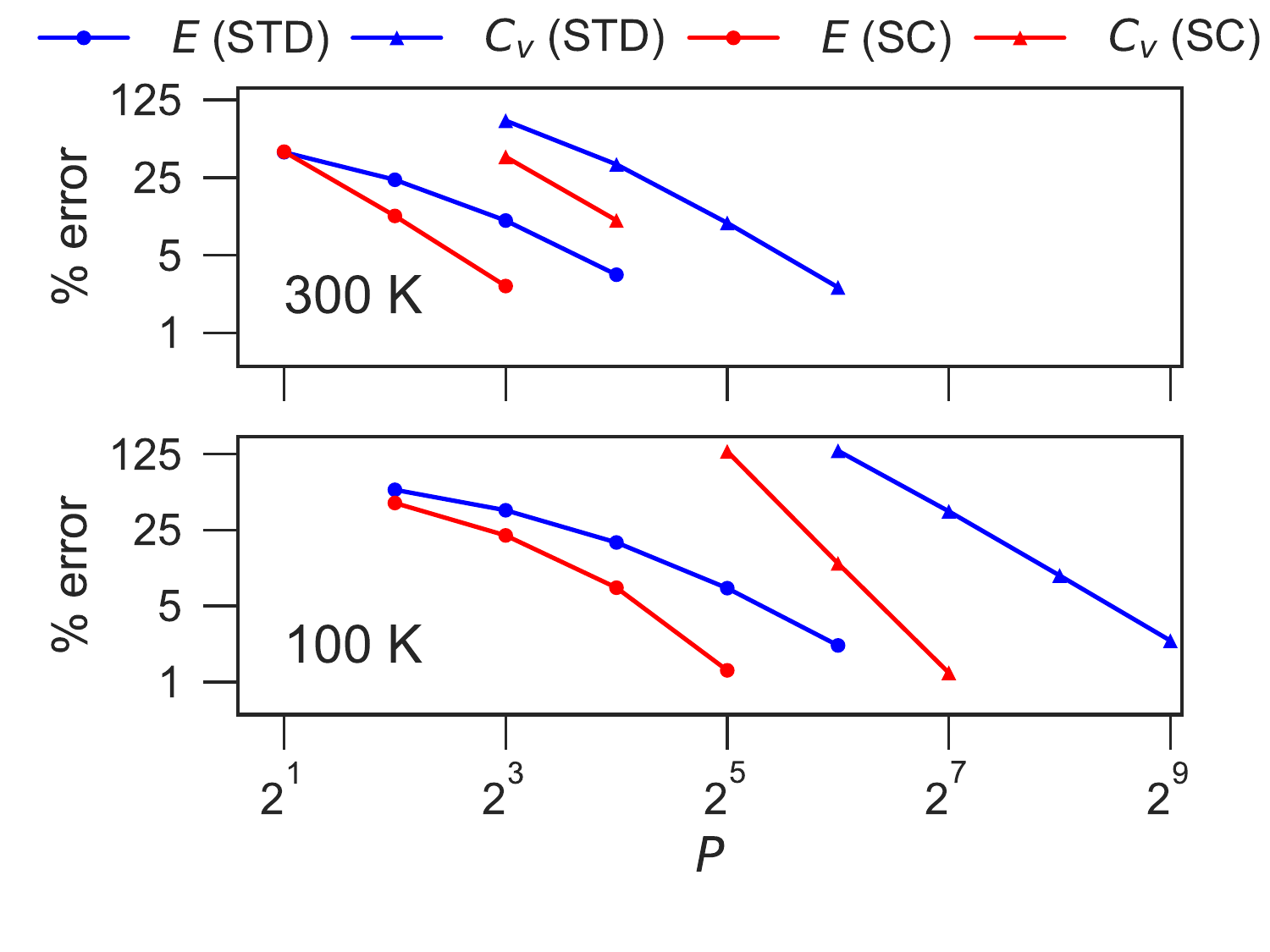}
\caption{Fractional error in the standard (red) and Suzuki-Chin (blue) path integral estimators of the energy (circles) and the heat capacity (triangles) of empty MOF-5 modeled by the corresponding Debye crystal potential, as a function of the number of beads $P$ at 100 K (bottom) and 300 K (top). The values of the energy and the heat capacity were obtained analytically \cite{brainthesis}.  
 \label{fig:tr_vs_sc}}
\end{figure}

The improved efficiency of the high order scheme is demonstrated by studying the convergence of the total energy and its temperature derivative for a harmonic model of MOF-5 -- obtained by computing the dynamical matrix associated with the first-principles-derived force field described in the previous section \ref{ss:ff} -- with respect to the number of replicas. The analysis is done analytically at two different temperatures: $T = 100$ K and $300$ K, as described in Ref.\ \cite{brainthesis}. As shown in Figure \ref{fig:tr_vs_sc}, the high order scheme improves the convergence of both the energy and the heat capacity by a factor of two at room temperature. In the low temperature or high accuracy regime where quantum effects dominate, the efficiency of the high order scheme is even more significant. 

The high accuracy afforded by the SC scheme makes it particularly useful to compute the heat capacity. 
This far, however, it has only been successfully applied to relatively simple systems \cite{jang+14jcp, pere-tuck11jcp,uhl+16jcp} and heat capacities in particular have only been reported for small clusters of molecules and constant-volume conditions. \cite{yama05jcp}
The difficulty in applying fourth-order schemes to complex materials can be understood by considering the fact that the force and virial contain derivatives of $\tilde{V}$ with respect to the atomic positions and the cell parameters,
\begin{equation}
\tilde{\vector{f}}^{(j)}\equiv -\frac{\partial \tilde{V}\left(\vector{q}^{(j)}\right)}{\partial \vector{q}^{(j)}}, \quad \tilde{\mathbf{\Xi}}^{(j)}\equiv\frac{\text{d} \tilde{V}
\left(\vector{q}^{(j)}\right)}{\text{d} \tensor{h}}\tensor{h}^{T} .
\end{equation}

Given that $\tilde{V}\left(\vector{q}^{(j)}\right)$ already contains first-order derivatives of the physical potential, the computation of the forces and the virial, required to sample the isothermal-isobaric ensemble by means of path integral dynamics, also demands the evaluation of higher-order derivatives of the potential, which is often cumbersome and computationally prohibitive.
Much of the work on the practical implementation of high order path integrals has therefore focused on avoiding the calculation of these terms, by sampling the standard path integral Hamiltonian and introducing fourth-order statistics by re-weighting \cite{jang-voth01jcp,pere-tuck11jcp}.
Unfortunately, re-weighting schemes have poor statistical performance for large systems \cite{jang-voth01jcp,ceri+12prsa}, so the application of the SC scheme has until now been limited to small systems and to constant-volume sampling.

To circumvent these limitations, we evaluate the virial using a finite-difference scheme that has recently been introduced by some of the present authors, to sample the SC canonical partition function directly. We discuss the essential ingredients of this scheme in Appendix \ref{aa:uvfx} and report a detailed derivation in the SI. Direct access to the instantaneous force and virial allows us to develop an integration scheme that samples the quantum isothermal-isobaric ensemble. This scheme can be regarded as an extension of the second-order scheme introduced in the pioneering work of Martyna and co-workers \cite{mart+99jcp}. The rather cumbersome derivation and the equations of motion are given in the SI. It should be noted that our implementation in \texttt{i-PI} \cite{ipi2} is also fully compatible with multiple time stepping \cite{tuck+92jcp,kapi+16jcp} and stochastic thermostatting, extending the integrators introduced in Ref.\ \cite{leim+13jcp} to the isothermal-isobaric ensemble. 

By using the finite-difference expressions, only $P/2$ additional force evaluations are needed instead of $P/2$ Hessians.  It is also useful to note that these components have a $P^{-2}$ prefactor, which means that they become small and slowly varying for typical values of $P$ used in SC simulations. \cite{kapi+16jcp2} This facilitates the use of a long time step for integrating the high order forces and virials, enabling us to sample the ensemble while evaluating the expensive terms rather infrequently. 

To further reduce the computational effort, we combine this scheme
with other accelerated PIMD methods that rely on the separation of the total potential into a cheap short-ranged term and an expensive long-ranged term, which is the case in our study, as discussed in subsection \ref{ss:ff}. We consider in particular ring polymer contraction (RPC) \cite{mark-mano08jcp} and multiple time stepping (MTS) \cite{tuck+92jcp}. Within the first method, the long-ranged components can be computed separately on a smaller ring polymer of $P'$ beads, which are subsequently, without loss in accuracy, extrapolated to the case of $P$ beads. The second method reduces the frequency at which the long-ranged interactions are computed by the use of a longer time step for the integration. The two methods can also be used together to achieve substantial computational savings \cite{kapi+16jcp,kapi+18jpcb} and can be seamlessly combined with our high order constant pressure scheme. 

\subsection{Calculation of thermodynamic observables}
Within the SC scheme, the thermodynamic averages of all structural observables $A$ are estimated by an ensemble average over the odd replicas:
\begin{align}
\mathcal{A}^{\text{OP}} =  \frac{2}{P}\avg{\sum_{j=1}^{P/2} A(\vector{q}^{(2j-1)})},
\end{align}
as first demonstrated by Jang and Voth \cite{jang-voth01jcp}. These estimators are commonly referred to as ``operator'' (OP) estimators, as opposed to the ``thermodynamic'' estimators obtained by derivation of the path integral partition function, that have often pathological statistical behavior. The simplicity of OP estimators makes the SC scheme very appealing in comparison to other high order schemes, in which \textit{ad hoc} estimators need to be constructed for simple structural observables \cite{pere-tuck11jcp,Poltavsky2018}. The simple OP estimators for the total energy $(\mathcal{E})$ and enthalpy $(\mathcal{H})$ are listed in Appendix \ref{aa:estimators}. Since the standard estimators for the heat capacity in path integral methods tend to be very complex and exhibit a large variance, we derive an OP double-virial estimator for the isobaric (and isochoric) heat capacity $C_P= \frac{\partial \mathcal{H}}{\partial T}$. The derivation is presented in the SI (Section S3.2) and the resulting expression is given in Appendix \ref{aa:estimators}, where it is also shown that this estimator has very good statistical properties and outperforms existing heat capacity estimators \cite{yamamoto2005}. 
However, in this study we computed thermophysical properties over a broad range of temperatures and found it more convenient to estimate $C_P$ by means of a finite difference approximation to the temperature derivative of the enthalpy:
\begin{align}
C_P(T) = \frac{\partial \mathcal{H}}{\partial T}  \approx   \frac{\left.\mathcal{H}\right|_{T + \Delta T} - \left.\mathcal{H}\right|_{T - \Delta T}}{2 \Delta T}.
\end{align}
A dedicated estimator will prove useful in simulations that are
targeted at a single, specific temperature.
\section{\label{sec:computational_details} COMPUTATIONAL DETAILS}

The required first-principles cluster data~\cite{tafipolsky2007} for the determination of the covalent terms in the force field are generated with \texttt{Gaussian 16}~\cite{gaussian16} using the B3LYP~\cite{b3lypa,b3lypb,b3lypc} exchange-correlation functional. A 6-311G($d$,$p$) basis set \cite{Poplebasisset} is used for the C, O, and H atoms, together with the LanL2DZ basis set for Zn. \cite{Lanl} The atomic charges are derived with the Minimal Basis Iterative Stockholder (MBIS) partitioning scheme \cite{verstraelen2016}. The atomic charges of the MOF-5 clusters are obtained from the PBE \cite{perdew1996} electron density computed with \texttt{GPAW} \cite{gpaw1, gpaw2} as implemented in \texttt{Horton} \cite{horton}. For methane, the atomic charges are derived from the B3LYP all-electron density obtained with \texttt{Gaussian 16}. The parameters of the van der Waals interactions are taken from the MM3 force field. \cite{lii1989,allinger1994}.
The van der Waals interactions are calculated up to a cutoff of 15 \AA\ and a tail correction is added to the potential and its derivatives.~\cite{tailcorr} The initial configurations of the methane molecules are generated using \texttt{RASPA} \cite{raspa} by inserting methane molecules at random positions, while ensuring that only realistic intermolecular distances are retained. Afterwards a canonical Monte Carlo algorithm was used to equilibrate the positions.    

For MOF-5 with and without methane, we perform classical and path integral MD simulations at a mechanical pressure of 1 bar and at different temperatures in the range of 100 K to 500 K. The classical MD simulations of both loaded and pristine MOF-5 are performed using \texttt{Yaff} in the $N\mathcal{P}(\tensor{\sigma}_a=\mathbf{0})T$ ensemble, \textit{i.e.}\ with no constraints on the unit cell. \cite{rogge2015} While the covalent interactions are calculated by \texttt{Yaff}, the expensive long-range interactions are computed by \texttt{lammps}~\cite{lammps} in a computationally efficient manner. The equations of motion are updated via a Verlet scheme, with a time step of 0.5 fs. The temperature is controlled via a single Nos\'{e}-Hoover chain consisting of three beads, with a relaxation time of 100 fs. \cite{nose1,nose2,nose3} A Martyna-Tobias-Klein barostat with a relaxation time of 1000 fs is used to control the pressure. \cite{mtk1,mtk2,rogge2015} We performed five independent runs of 500 ps, starting from a different random seed and from different methane positions. For the empty MOF-5, a single trajectory of 500 ps was used. An equilibration time of 100 ps was considered. 

The PIMD simulations are performed with the universal force engine \texttt{i-PI} \cite{ipi1,ipi2} in the $N\mathcal{P}(\tensor{h}_{0})T$ ensemble, where the cubic symmetry is kept fixed. The evaluation of the forces is carried out by \texttt{Yaff} and \texttt{lammps}, similar to the classical MD simulations, while the time evolution of the nuclei to sample the appropriate thermodynamic ensemble is done with \texttt{i-PI}. To control the temperature, a PILE-L thermostat \cite{ceri+10jcp} is applied to the system and a white noise Langevin thermostat \cite{buss-parr07pre} is applied to the cell. To control the pressure, a path-integral version of the Bussi-Zykova-Parinello (BZP) barostat \cite{bzp,ipi1}, adapted to the SC scheme, is used. The time constants for the thermostats and the barostats are same as the ones used in the classical simulations. A $BAOAB$ type \cite{leim+13jcp} MTS scheme \cite{tuck+92jcp} (see SI Section S2.3) is used to integrate the equations of motion. The computationally cheap short-range terms of the force field are computed on 64 replicas and integrated with a time step of 0.25 fs. The remainder of the interactions, \textit{i.e.}\ the expensive long-range interactions, are computed on $8$ replicas using RPC and integrated with a time step of 1 fs. As discussed above, a finite differences strategy is adopted to determine the heat capacity from the enthalpy with a temperature interval of 25 K. We performed thirty independent runs of 50 ps, starting from a different random seed and from different methane positions.  For the empty MOF-5, five independent trajectories of 125 ps were used. An equilibration time of 25 ps was considered.

\begin{figure*}[!t]
	\begin{center}
	\includegraphics[width=\linewidth]{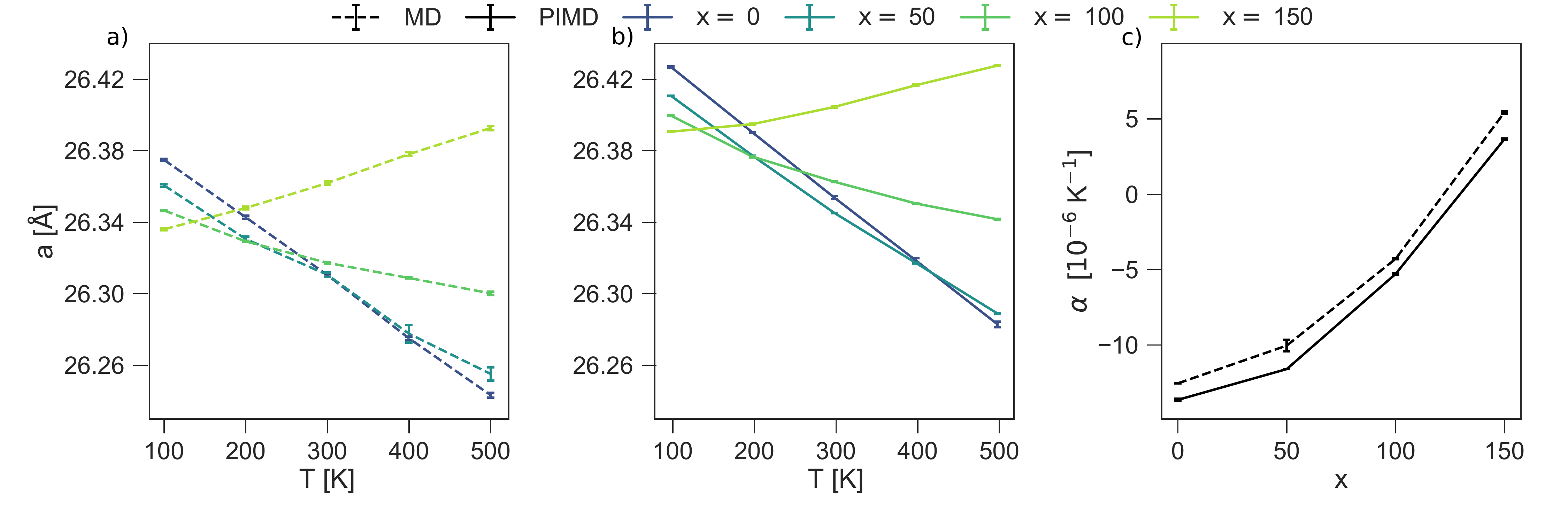}
	\end{center}
	\caption{\label{fig:mof_volume} Panels (a) and (b) show the lattice parameter $a$ of MOF-5 with $x=0, 50,100,150$ molecules of methane as a function of temperature ($T$), obtained from classical MD and PIMD respectively. Panel  (c) shows the linear thermal expansion coefficient ($\alpha$) as a function of $x$. The classical and quantum estimates are respectively shown with dashed and solid lines. Error bars indicate statistical uncertainity.}  
\end{figure*}

\section{\label{sec:results} Results}

Having developed an accelerated integration scheme for the quantum isothermal-isobaric ensemble, we now focus on the structural and thermal properties  of methane-loaded MOF-5. Extensions towards other MOFs and adsorbates will be the topic of future studies. The importance of the inclusion of NQEs and anharmonicities in the modeling of the heat capacity is probed by comparing the results with other methods such as classical MD, which neglects NQEs, and the harmonic approximation, which neglects anharmonicity. We discuss the accuracy of these commonly-adopted approximations and provide empirical relations, which might resolve the general lack of knowledge on the heat capacity of this class of materials.

\subsection{Structural properties}

To unravel the influence of adsorbates on the framework and finally on the heat capacity, we start by investigating the structural response of MOF-5 for various loadings and temperatures. Here, one could expect that a proper inclusion of NQEs already becomes important as zero-point effects were recently found to substantially increase the volume of MOF-5 when comparing classical MD with PIMD. \cite{lamaire2018} Additionally, NQEs have previously been observed to change the volume of bulk alkanes by about $10 \%$. \cite{Balog2000,Pereyaslavets2018, veit2018} A comparison of Figures \ref{fig:mof_volume} (a) and (b) indeed reveals that the inclusion of NQEs increases the volume by almost 1 \% for all loadings and temperatures. Horizontally, this shift corresponds to a more substantial temperature reduction of about 100 K.

Interestingly, the qualitative ordering of the volume as a function of loading at the different temperatures does not change appreciably with or without the inclusion of NQEs. At low temperatures, the material slightly shrinks in the presence of methane. The observed adsorption-induced deformation can be understood by attractive van der Waals interactions between the framework and the adsorbed methane. \cite{ravikovitch2006,joo2013} At higher temperatures, by contrast, the empty framework has the lowest volume, as entropic and kinetic effects start dominating and the adsorbed molecules increase the internal pressure, which leads to a volumetric expansion when increasing the loading.
The main effect of NQEs on the volume of the guest-loaded system is thus an upward volume shift, which is to a large extent independent of the number of guest molecules and the temperature, and thus originates from the zero-point fluctuations of the framework lattice. \cite{lamaire2018}

Varying the concentration and the type of adsorbates in the framework was suggested by Calero and co-workers \cite{balestra2016} as a way to control and tune the thermal expansion of a system based on classical MD simulations. We confirm that with methane, it is possible to go from the well-known negative thermal expansion behavior of the empty framework \cite{zhou2008,lock2010,lock2013} towards positive thermal expansion. A proper inclusion of NQEs in our molecular dynamics simulations does not influence the predictions in this temperature window. The role of the quantum mechanical nature of the framework nuclei also remains limited when studying the thermal expansion coefficient, as both classical MD and PIMD simulations lie within the experimental range. \cite{lamaire2018} This conclusion does not change with methane in the pores, as shown in Figure \ref{fig:mof_volume} (c). 

\begin{figure*}[htp]
	\begin{center}
		\includegraphics[width=\linewidth]{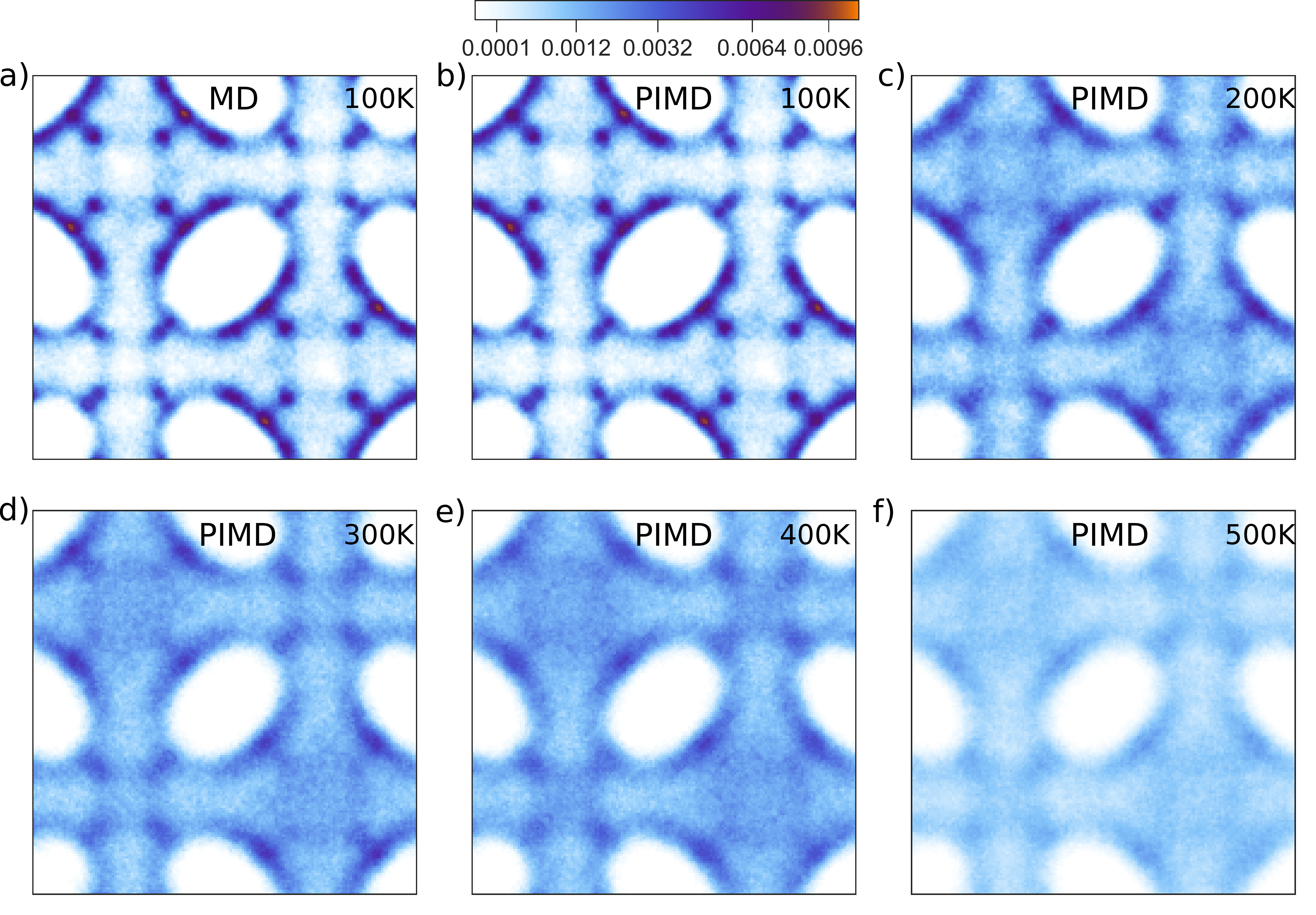}
	\end{center}
	\caption{\label{fig:mof_loadings} The methane distribution in the pores of MOF-5 at different temperatures as obtained from PIMD simulations. Orange spots indicate high probability adsorption sites. Other colors show the distribution of the low probability methane positions in the conventional unit cell and represent the probability representation (from very high (orange), to high (dark blue), to low (white) probability).}
\end{figure*}

A more surprising picture emerges when looking at the distribution of methane inside MOF-5. Recent PIMD simulations \cite{Pereyaslavets2018} of bulk methane (at 110 K) have shown that NQEs lead to significant changes in the structure of methane at low temperature, corresponding to an overall softening of the structure and an increase in the intermolecular distance by about 0.1 \AA. In contrast, in our study of methane confined in the pores of MOF-5, even at 100 K -- where NQEs are expected to be the greatest -- there is no appreciable difference between the shape of the classical and quantum distribution functions of methane, as shown in Figure \ref{fig:mof_loadings}. This can be understood by the fact that the change in the structure of bulk methane comes entirely from the isotropic expansion of the gas \cite{note-1}. In the case of  methane molecules confined in the pores,  the low compressibility  of the framework makes the expansion as observed in bulk methane when including NQEs impossible.

This discussion shows that the structural response of MOF-5 to a varying number of methane molecules and temperature is largely unaffected by NQEs, except for the zero-point lattice fluctuations. Our observations also corroborate the common practice of ignoring NQEs when studying the loading of porous materials by Grand Canonical Monte Carlo simulations \cite{raspa}.  Nevertheless, this conclusion cannot be generalized to other adsorbates, especially those possessing stronger intermolecular interactions such as hydrogen bonding. 

\begin{figure}[htp]
	\begin{center}
		\includegraphics[width=0.50\linewidth]{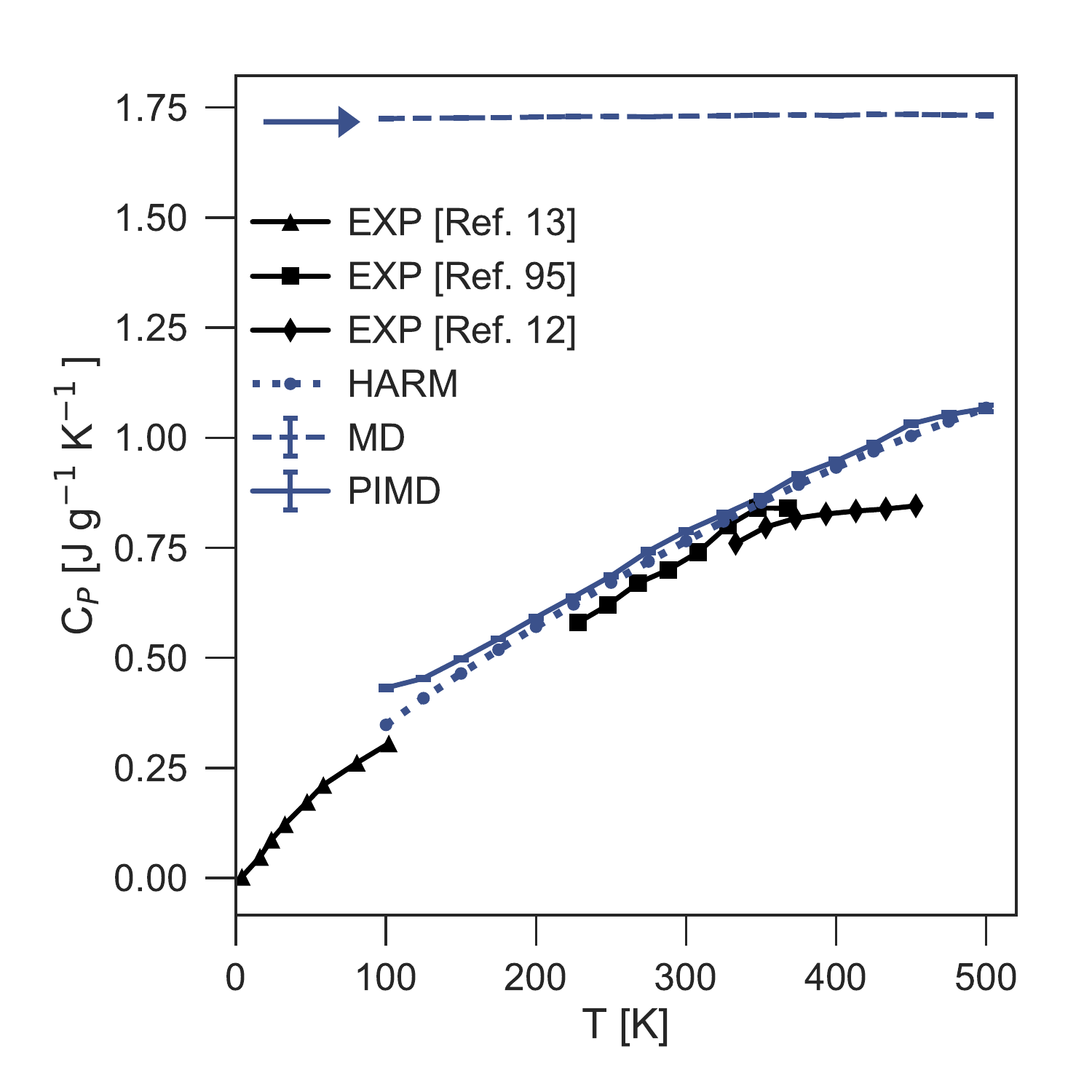}
	\end{center}
	\caption{\label{fig:mof_clean} Heat capacity $C_P$ of the empty MOF-5 as a function of temperature $(T)$ computed using classical MD (dashed), PIMD (solid) and the harmonic  approximation (dotted). The right pointing arrow shows the Dulong-Petit limit. Different experimental results are shown in black using triangular \cite{kloutse2015}, square \cite{ming2014} and diamond \cite{mu2011} markers. Error bars indicate statistical uncertainity. }  
\end{figure}

\subsection{Heat Capacity}

MOF-5 has been the subject of a few experimental heat capacity studies \cite{mu2011,liu2012,ming2014,kloutse2015} which have shown that the material has a low specific (or molar) heat capacity, about 0.7 J/g$\cdot$K at room temperature, even when compared to other MOFs. Depending on the type of the application, a large (\textit{e.g.}\ for ANG to limit temperature fluctuations) or a small (\textit{e.g.}\ for CCS to limit the energy penalty) heat capacity is sought after. It is thus important to understand how this property changes at different levels of loading and temperature, and to determine the factors influencing the heat capacity, which is now possible for the first time using our high order PIMD scheme. 

In the previous section, it has been shown that classical MD can -- at least qualitatively -- be used to model the structural response of MOF-5 in the presence of methane at various temperatures. This approach is however expected to fail for the description of the heat capacity since the heat capacity of many systems is dominated by NQEs at room temperature, as evidenced by experimental deviations from the classical Dulong-Petit law. The most common way of including NQEs for solids is the static harmonic approximation, using Einstein’s or Debye’s harmonic model for solids, which is able to reproduce the heat capacity of many solids and will therefore also be used for comparison. 

We begin by presenting the estimates of the temperature dependence of the isobaric heat capacity of the empty MOF-5 framework. As shown in Figure \ref{fig:mof_clean}, the classical MD estimates (dashed line) are in agreement with the Dulong-Petit law. The simulations yield an almost constant value of $3$ k$_B$ per degree of freedom, which indeed results in large deviations from the experimental values. \cite{mu2011,liu2012,ming2014,kloutse2015} Upon inclusion of NQEs with our PIMD scheme (solid line), we find that the results follow the experimental measurements reasonably well up to almost 400 K. This agreement is remarkable as these measurements are typically carried out on the as-synthesized sample, which possibly includes solvents \cite{mu2011} and differs from the perfect crystal that we have simulated. 
\begin{figure*}[htp]
	\begin{center}
	\includegraphics[width=\linewidth]{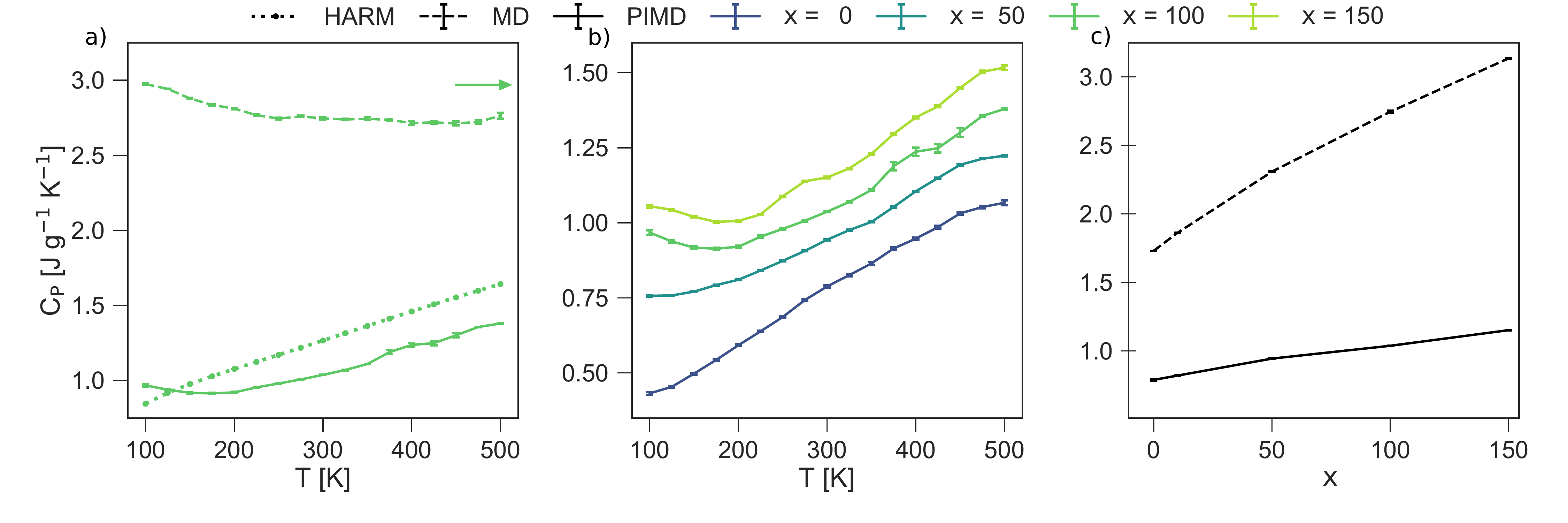}
	\end{center}
	\caption{\label{fig:fig_MOF_cp} Panel (a) shows the comparison of the classical (dashed), quantum (solid), and harmonic estimates (dotted) of the isobaric heat capacity $C_P$ of MOF-5 with 100 molecules of methane, as a function of the temperature $T$. Panel (b) shows the temperature dependence of the quantum isobaric heat capacity of MOF-5 with $x$ molecules of methane. Panel (c) shows the quantum isobaric heat capacity of the MOF with $x$ molecules of methane as a function of $x$ at 300 K. Error bars indicate statistical uncertainity. }  
\end{figure*}
Figure \ref{fig:mof_clean} also reveals that the results obtained using the simple and computationally inexpensive harmonic approximation (dotted line) are in good agreement with the exact values computed with PIMD. This implies that anharmonic quantum contributions to the heat capacity and the effect of an adequate anharmonic sampling are small for the empty MOF. Moreover, the harmonic approximation using the  UFF4MOF force field \cite{addicoat2014} yields largely similar results (see SI  Section S6.1), so that a less accurate inexpensive and generic model for the potential energy surface is capable of reproducing the heat capacity of the empty framework. 

Another notable detail of our calculations is that the harmonic approximation was used to estimate the isochoric heat capacity instead of the isobaric one. As the isobaric and isochoric heat capacities are almost the same, the MOF behaves like a regular solid, despite its large negative thermal expansion coefficient. The harmonic approximation could therefore serve as an efficient procedure to accurately estimate the heat capacity of the empty framework in the increasing number of high-throughput MOF screenings. \cite{wilmer2012,huck2014,simon2015,moghadam2018,tabor2018} 

In order to study the effect of adsorbates, we start by considering the case of a loading $x$ of 100 methane molecules per conventional unit cell (8(Zn$_4$O(CO$_2$)$_6$) $\cdot$ 100 CH$_4$). Although a high-level PIMD strategy might not be required to estimate the heat capacity of the empty MOF host, PIMD proves to be crucial to capture the correct temperature dependence of the loaded system, as can be seen in the left panel of Figure \ref{fig:fig_MOF_cp}. Here, anharmonic effects become important as we observe differences between the PIMD results and the harmonic approximation. The discrepancy in the qualitative behavior of the heat capacity between both techniques can be understood through the mobility of the guest molecules in the large pores of the framework, which cannot be adequately captured by a harmonic approximation. \cite{tsivion2017} These low-frequency anharmonic motions explain why we find at the classical MD level a similar low-temperature dependence as in PIMD, but only PIMD simulations include both anharmonic and nuclear quantum effects correctly. Interestingly, the combination of both effects yields a heat capacity that does not change monotonically, but exhibits a minimum at about 200 K.

\begin{figure*}[t]
\begin{center}
		\includegraphics[width=\linewidth]{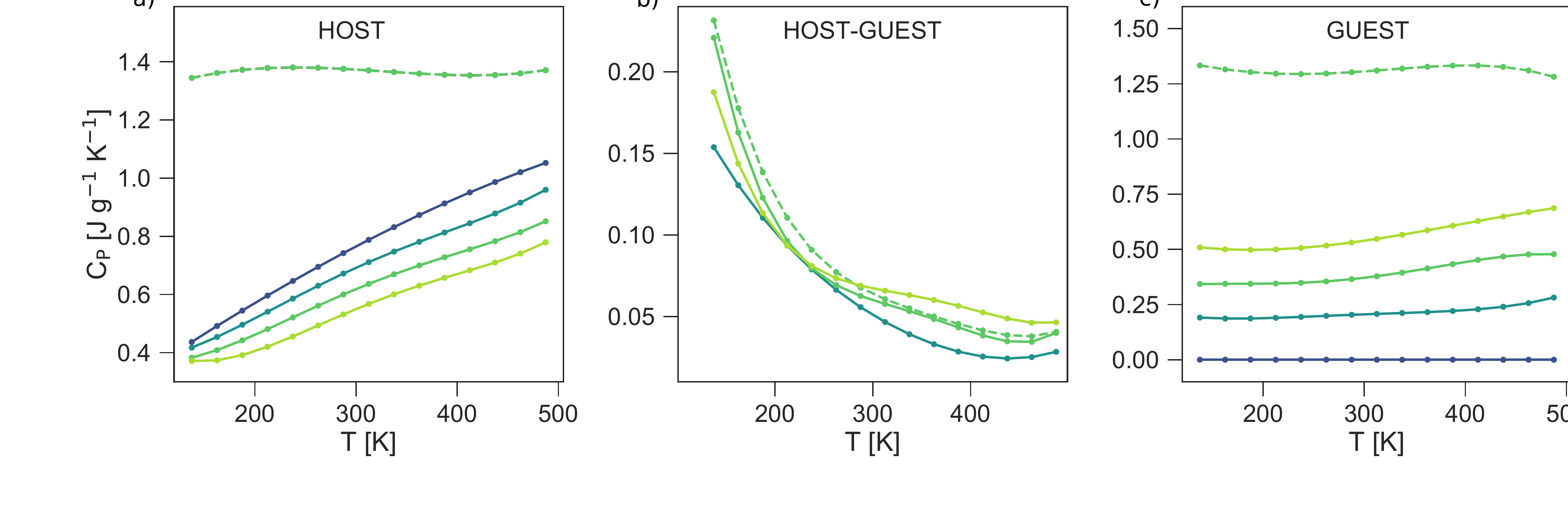}
\end{center}
\caption{\label{fig:fig_decomp} Panels (a), (b) and (c) respectively show the decomposition of the specific heat capacity of the MOF and the adsorbate system into host, host-guest, and guest-guest contributions for different gas loadings $(x)$.  The curves were obtained by deriving a polynomial fit to the energy as a function of temperature.}
\end{figure*}

Extending towards other loadings of methane in the middle panel of Figure \ref{fig:fig_MOF_cp}, it becomes clear that the heat-capacity minimum as a function of temperature depends strongly on the number of guests and becomes more pronounced at higher loadings. Even when expressing the heat capacity normalized to the total mass of the system, one can see that at a fixed temperature $C_p$ increases almost linearly with the loading (see Figure \ref{fig:fig_MOF_cp} (c) at 300 K). For the volumetric heat capacity, \textit{i.e.}, the heat capacity per unit of volume of the system, similar results are obtained (see SI Section S6.3).  

\begin{figure}[t]
\begin{center}
		\includegraphics[width=0.5\linewidth]{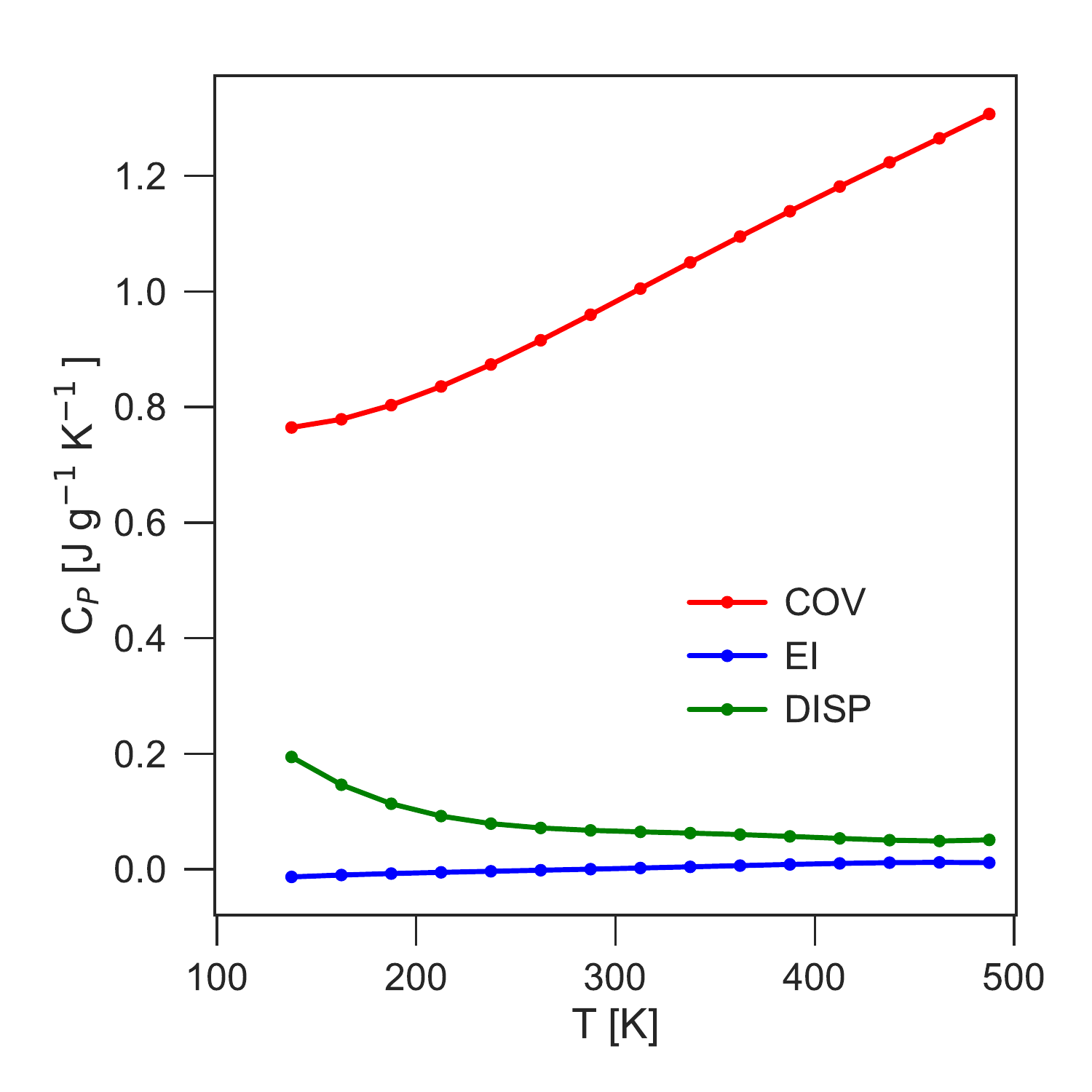}
\end{center}
\caption{\label{fig:fig_decomp_ii} The decomposition of the total heat capacity of MOF-5 with 100 methane molecules per unit cell into covalent (COV), electrostatic (EI), and van der Waals (DISP) contributions. The curves were obtained by deriving a polynomial fit to the energy as a function of temperature.}
\end{figure}

To rationalize the origin of the non-monotonic temperature dependence of the heat capacity, we determine which interactions give the most substantial contribution to $C_p$. To this end, the force-field energy contributions are decomposed in terms of the host, host-guest, and guest-guest interactions (see SI Section S6.2). Figure \ref{fig:fig_decomp} displays the most important results of this analysis. The host and guest-guest contributions to the specific heat capacity are visualized in panels (a) and (c). The shape of the different host curves appears to be independent of the loading. In fact, when rescaled to the mass of the empty MOF, the curves coincide with one another and with the curve obtained within the harmonic approximation. This demonstrates that the degrees of freedom of the MOF-5 framework, which are more strongly quantized, are predominantly harmonic and do not significantly change due to the interaction with methane. Their contribution to the total heat capacity per unit mass, however, decreases with the loading due to a change in the mass balance. The guest-guest interactions, on the other hand, are relatively constant and only show a small increase when going from 100 K to 500 K, due to the activation of high-frequency vibrational modes. The most interesting contribution arises from the host-guest interactions, which explains the non-monotonic behavior of the specific heat capacity of the guest-loaded system. The contribution of these interactions decreases with a sharp temperature dependence when sufficient guest molecules are present inside the pores. The large heat capacity at temperatures lower than 100 K originates from the known first-order structural phase transition of methane in MOF-5 at 60 K, \cite{wu2009,kuchta2017} from which we observe the decreasing tail. Since the methane molecules are more localized at low temperatures, the attractive host-guest interactions allow to efficiently store thermal energy. At higher temperatures, from 250 K to 500 K, the host-guest contributions become negligible as the confined guests become more mobile and less bound to the framework, so that the increase in thermal energy can no longer be stored in the physical interactions between the methane guests and the MOF-5 host.

Another decomposition of the force-field energy in terms of the covalent, electrostatic, and van der Waals interactions shows that the short-range covalent interactions and thus the network of chemical bonds (Figure \ref{fig:fig_decomp_ii}) dominate the contributions to the heat capacity. For the empty MOF-5 framework, the noncovalent interactions are negligble (see SI Section S6.2). This confirms that the heat capacity of empty MOF-5 can be approximated by considering only contributions from the separate molecular fragments of the material \cite{mu2011} and suggests why the harmonic approximation works well for this material. For the loaded framework, the noncovalent part starts to play a role, which is especially true for the host-guest interactions. Not surprisingly, in the case of nonpolar methane molecules, these interactions are dominated by the van der Waals terms in the force field (see SI Section S6.2). This suggests that the use of different, more polar, guests in which electrostatic interactions play a more prominent role (\textit{e.g.}\ CO$_2$) could give rise to other interesting phenomena. However, care must be taken in interpreting these different terms as a separation is not unambiguously defined and might be force-field dependent.

\subsection{The interplay of gas loading, anharmonicities, and quantum effects}

Our analysis of the structural and thermal properties of methane-loaded MOF-5 shows that the total system does not always need a full treatment of anharmonicities and NQEs. This suggests that a full path integral sampling of the entire system may not be necessary, especially if qualitative trends are to be studied. Hence, inspired by our results, we propose an empirical formula for the volume and the heat capacity in which the most important effects, \textit{i.e.}, anharmonicities and/or NQEs, are captured and which might prove to be beneficial for future studies of guest-loaded MOFs.

As discussed above, the main difference between the volume with or without NQEs comes from zero-point effects in the lattice. The correct volume can therefore be estimated as follows: 
\begin{align*}
     \mathcal{V} \approx \mathcal{V}^{\text{anh}}_{\text{qn}}[\text{MOF-5}] - \mathcal{V}^{\text{anh}}_{\text{cl}}[\text{MOF-5}] + \mathcal{V}^{\text{anh}}_{\text{cl}}[\text{MOF-5} + \text{CH}_{4}],
\end{align*}

where anh stands for the inclusion of anharmonicities with MD, and cl and qn denote the use of classical or path integral MD respectively. The left most panel of Figure \ref{fig:mof_approx} indicates that this approximate volume agrees very well with the exact results obtained from PIMD simulations. A more stringent test is the thermal expansion coefficient, which is -- as shown in Figure \ref{fig:mof_approx} (b) -- also in excellent agreement with the PIMD results. For systems where a first-principles treatment of the potential energy surface is required and PIMD simulations are too expensive, other approximate techniques such as the quasi-harmonic approximation or classical MD with a quantum thermostat \cite{ceri+09prl} could be used to estimate the zero-point effects \cite{lamaire2018}. 

In contrast, we observed in the previous section that the heat capacity of the framework could be estimated with a harmonic approximation, while the guest-host interactions are dominated by anharmonicities. For that reason, we propose: 
\begin{align*}
     {C} \approx \left(C^{\text{har}}_{\text{qn}} - {C}^{\text{har}}_{\text{cl}} + {C}^{\text{anh}}_{\text{cl}}\right)[\text{MOF-5} + \text{CH}_{4}],
\end{align*}
in which the high frequency modes of adsorbate and the MOF are treated in a harmonic fashion and the host-guest interactions are treated classically. For ${C}^{\text{har}}_{\text{cl}}$, the Dulong-Petit law can be used. As shown in the rightmost panel of Figure \ref{fig:mof_approx}, our suggested approximation is even able to qualitatively reproduce the heat capacity minimum for a loading of $100$ methane molecules. Beyond 200 K, the agreement between the empirical expression and the exact PIMD becomes quantitatively correct. This method could thus be an inexpensive route to estimate the heat capacity of guest-loaded MOFs. 

\begin{figure*}[t]
	\begin{center}
		\includegraphics[width=\linewidth]{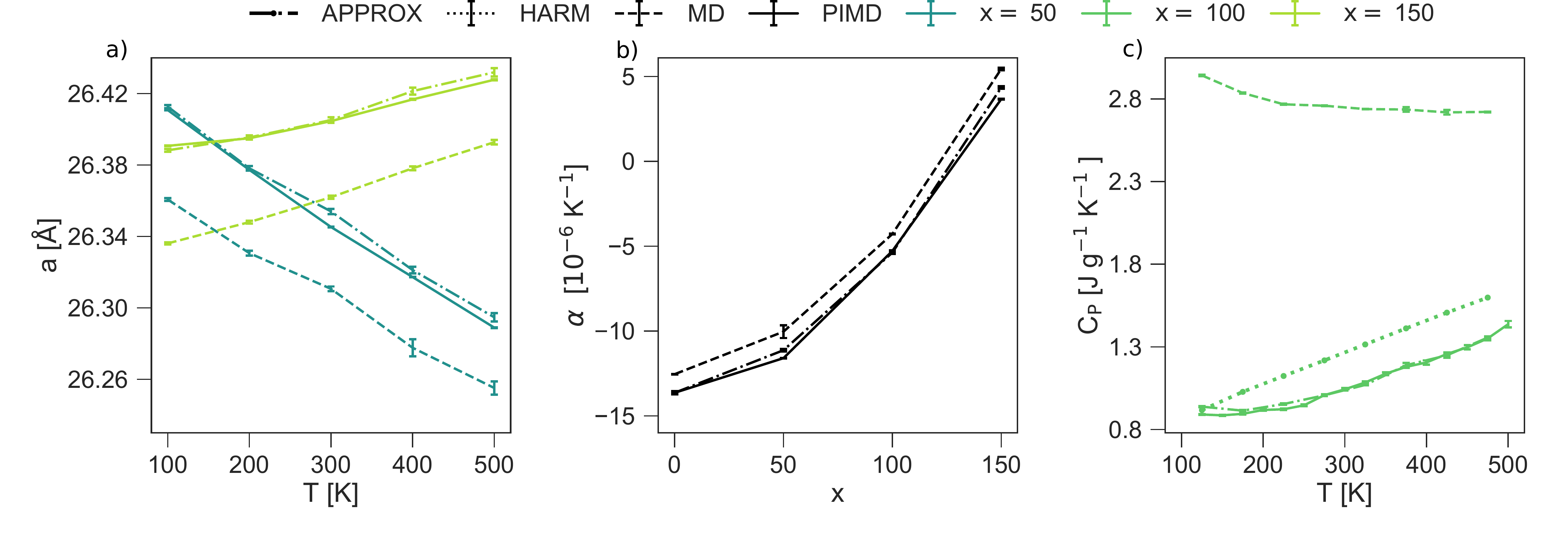}
	\end{center}
	\caption{\label{fig:mof_approx} Panels (a), (b) and (c) respectively show the temperature dependence of the cell parameter ($a$) for MOF-5 with 50 and 150 molecules of methane, the linear thermal expansion coefficient $\alpha$ of MOF-5 with $x$ molecules of methane as function of $x$, and the isobaric heat capacity $C_P$ of MOF-5 with 100 molecules of methane, obtained with classical MD (dashed),  PIMD (solid), and the approximation introduced in the work (dot-dashed). Error bars indicate statistical uncertainity. }
\end{figure*}

\section{Conclusions}

To summarize, we developed an efficient and accurate methodology to calculate the isobaric thermophysical properties of materials, that is generally applicable and therefore ideally suited to the study of guest-loaded MOFs. For this purpose, we derived and implemented the necessary algorithms in \texttt{i-PI} \cite{ipi2} to perform simulations in the quantum isothermal-isobaric ensemble using the Suzuki-Chin path integral molecular dynamics framework. The method is rigorous and can be seamlessly combined with other cost-reduction techniques, which facilitates a huge reduction of the computational cost compared to standard techniques. 

We demonstrated the applicability of our approach by investigating the heat capacity of the prototypical MOF-5 loaded with different numbers of methane molecules. We observed that the level of statistical sampling that is needed to achieve quantitative accuracy depends on the property of interest. 
For all the cases we considered, we found the framework to behave in a strongly quantized manner, but to be largely amenable to a harmonic treatment. The adsorbates, on the other hand, show only mild quantum effects in their intermolecular interactions, but require a full anharmonic description. 
The heat capacity shows a particularly subtle interplay of quantum and anharmonic fluctuations, that results in a non-monotonic temperature dependence of the heat capacity with a minimum. 

Through a decomposition of the heat capacity into molecular interactions, we find that the host-guest interactions are responsible for this behavior, as their contribution to the total heat capacity decreases with temperature. 
By comparing the behavior of different classes of framework materials and guest molecules, this may reveal new design rules to optimize the thermal behavior of a storage material over a broad range of temperatures and levels of loading. 
Our approach provides an affordable route to perform benchmark studies and approximation strategies to carry out the high-throughput studies that are needed to obtain a complete understanding of the interplay between framework, adsorbate, and quantum mechanical and anharmonic fluctuations that determine the thermophysical properties of MOFs.

\section{Acknowledgements}

We  acknowledge  financial  support  by  the  Swiss  National  Science  Foundation  (project  ID  200021-159896), the Fund for Scientific Research Flanders (FWO), the Research Board of Ghent University and the European Union's Horizon 2020 research and innovation programme (consolidator ERC grant agreement No. 647755-DYNPOR (2015-2020)). The computational resources and services used in this work were provided by VSC (Flemish Supercomputer Center), funded by Ghent University, FWO, and the Flemish Government department EWI. We also want to acknowledge R.\ Semino, S.M.J.\ Rogge and L.\ Vanduyfhuys for their valuable input and discussions.

\section{Supporting Information}

Derivation of the Suzuki-Chin partition functions, forces, virials and total stress; the equations of motion and integration schemes for sampling the isothermal-isobaric scheme ensemble; derivation of the heat capacity estimators; details of the generation and validation of the first principles force field; temperature dependence of the adsorption isotherms for methane; the formula of the harmonic heat capacity, the decomposition of the total heat capacity into contributions from different physical interactions and the temperature dependence of the volumetric heat capacity. 

\section{Author Contributions}

VK and JW contributed equally to this work.

\appendix

\section{\label{aa:uvfx} Suzuki-Chin forces and virials}
Recently, some of the present authors showed that the high-order force can be estimated without computing the Hessian in a finite difference fashion \cite{kapi+16jcp2} (see SI Section S1.5):
\begin{align}
\begin{split}
\tilde{\vector f}^{(2j)} 
& = \omega_P^{-2} m^{-1} \frac{\partial}{\partial \vector{q}^{(2j)}} | \vector{f}^{(2j)}|^2 \\
& = -2\omega_P^{-2} \lim_{\epsilon \to 0}\frac{1}{\epsilon\delta}\left(\left.\vector{f}^{(2j)}\right|_{\vector{q}^{(2j)} + \epsilon\delta \vector{u}^{(2j)}} - \vector{f}^{(2j)}\right),
\end{split}
\end{align}
where $\vector{u}^{(2j)}=\vector{f}^{(2j)} / m$ and  $\delta = \left[(3NP)^{-1} \sum_{j=1}^{P} \vector{u}^{(j)}\cdot \vector{u}^{(j)}\right]^{-\frac{1}{2}}$ is a normalization factor, so that $\epsilon$ represents the root mean square displacement applied to each atom. This avoids the explicit calculation of the Hessian and allows for the direct sampling of the Suzuki-Chin canonical ensemble. Following a similar strategy (see SI Section S1.6), we show that the high-order component of the virial can be estimated as:
\begin{align}
\begin{split}
  \tilde{\mathbf{\Xi}}^{(2j)} & =  \left[- \sum_{i=1}^{N} \vector{q}_{i}^{(2j)} \otimes \tilde{\vector{f}}_{i}^{(2j)} + 
\frac{\partial \tilde{V}\left(\vector{q}^{(2j)}\right)}{\partial \tensor{h}} \tensor{h}^{T}\right] \\
& =  -2\omega_P^{-2} \left[\sum_{i=1}^N \vector{f}_i^{(2j)} \otimes \vector{f}_i^{(2j)}  / m \right.\\
& ~~\left. +  \lim_{\epsilon \to 0}\frac{1}{\epsilon\delta}\left(\left.\vector{\Xi}^{(2j)}\right|_{\vector{q} + \epsilon\delta \vector{u}^{(2j)}}  - \vector{\Xi}^{(2j)}\right)\right].
\end{split}
\end{align}
\section{\label{aa:estimators} Suzuki-Chin Estimators}

\begin{figure}[htp]
	\begin{center}
		\includegraphics[width=0.5\linewidth]{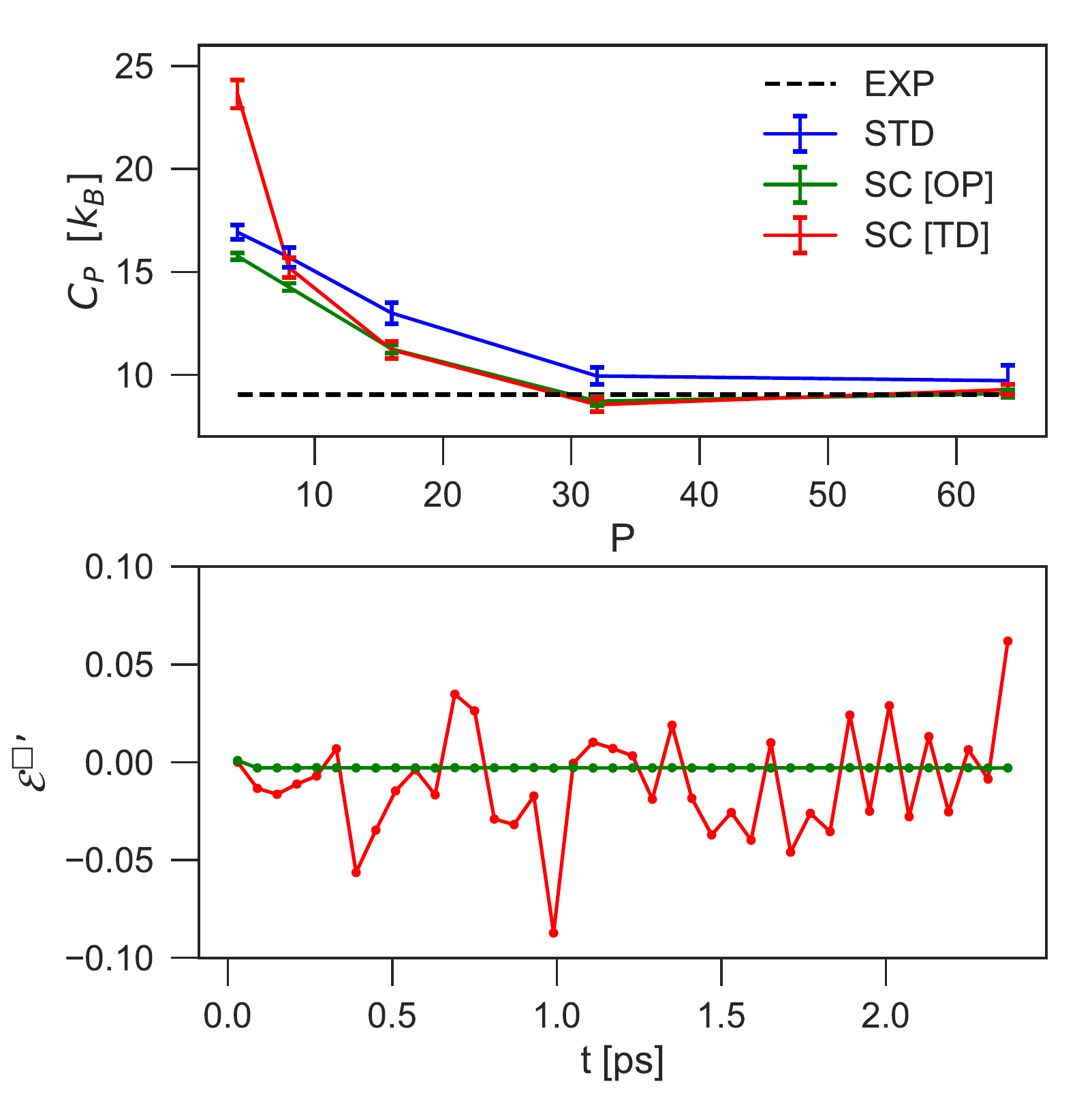}
	\end{center}
	\caption{\label{fig:est} The top panel shows the convergence of the isobaric heat capacity of liquid water at 300 K, modeled by the q-TIP4P/f potential, as a function of number of replicas. The blue curve was obtained from standard PIMD using the Yamamoto estimator while the red and green curves were obtained from SC PIMD using the Yamamoto (TD) and OP estimator respectively. The dashed black line represents the experimental result. Error bars indicate statistical uncertainity.  The bottom panel shows that instantaneous values of the the computationally expensive  ${\mathcal{E}^{\square}}'$ term when computed with the OP and the Yamamoto (TD) estimator. }  
\end{figure}

The operator (OP) and thermodynamic (TD) estimators of the energy for Suzuki-Chin isothermal-isobaric PIMD simulations of a single particle system, described in Section \ref{ss:acpimd}, are given by
\begin{align}
     \mathcal{E}^{\text{OP}} & = \frac{3}{2\beta} + \frac{2}{P} \sum_{j=1}^{P/2} \avg{ -\frac{1}{2}\left(\vector{q}^{(2j-1)} - \bar{\vector{q}}\right) \cdot \vector{f}^{(2j-1)}} \nonumber \\
     & ~+ \frac{2}{P} \sum_{j=1}^{P/2} \avg{V(\vector{q}^{(2j-1)}) } \\
     \mathcal{E}^{\text{TD}} & =  \frac{3}{2\beta}  + \frac{1}{P} \sum_{j=1}^{P} \avg{-\frac{1}{2}\left(\vector{q}^{(j)} - \bar{\vector{q}}\right) \cdot \vector{f}^{\text{sc}~(j)}} \nonumber \\ 
     & ~+ \frac{1}{P} \sum_{j=1}^{P/2} \avg{\frac{3}{9} \tilde{V}\left(\vector{q}^{(2j)}\right)} 
     + \frac{1}{P} \sum_{j=1}^{P} \avg{w_j V\left(\vector{q}^{(j)}\right)}
\end{align}
where $\bar{\vector{q}} \equiv \frac{1}{P} \sum_{j=1}^{P}\vector{q}^{(j)}$ is the position of the  centroid, $\vector{f}^{\text{sc}~(j)} \equiv - {\partial V^{\text{sc}}\left(\vector{q}^{(j)}\right)} / {\partial \vector{q}^{(j)}}$ the total Suzuki-Chin force, and $\avg{\cdot}$ denotes a thermodynamic average over the quantum isothermal-isobaric ensemble. The generalization to a many particle system is straightforward. The $\square =$ OP or $\square =$ TD estimators of the enthalpy are obtained through the simple expression:
 \begin{align}
     \mathcal{H}^{\square} & = \mathcal{E}^{\square} + \mathcal{P} \avg{\mathcal{V}}.
 \end{align}
 
The $\square =$ OP or $\square =$ TD (that we will refer to as the Yamamoto estimator~\cite{yamamoto2005}) double virial estimators of the heat capacity take the following form:
\begin{align}
    \frac{C_P^{\square}}{k_B \beta^2} = & \, \text{Cov}[\mathcal{H}^{\square}, \mathcal{H}^{\text{TD}}]  +  {\mathcal{E}^{\square}}',
\end{align}
where ${\mathcal{E}^{\square}}'$ for $\square =$ TD is given in Ref.\ \cite{yama05jcp} and 
\begin{align}
    {\mathcal{E}^{\text{OP}}}' = & \frac{3N}{2\beta^2} + \frac{1}{\beta P} \sum_{j=1}^{P/2} \avg{\left(\vector{q}^{(2j-1)} - \bar{\vector{q}}\right) \cdot \vector{f}_{\text{dcv}}^{(2j-1)}}.
\end{align}
The double virial force $\vector{f}_{\text{dcv}}^{(2j-1)}$, which depends on the Hessian of the physical potential, can be estimated in a computationally efficient way by finite differences:
\begin{align}
    \vector{f}_{\text{dcv}}^{(j)} = \frac{3}{2} \vector{f}^{(j)} + \lim_{\epsilon \to 0} \frac{1}{2\epsilon} \left[\left.\vector{f}^{(j)}\right|_{\vector{q}^{(j)} + \epsilon \left(\vector{q}^{(j)} - \bar{\vector{q}}\right)} - \vector{f}^{(j)}\right].
\end{align}
We test this estimator on the well known isobaric heat capacity of liquid water which is 1 $\text{cal} ~\text{mol}^{-1} K^{-1}$ or 9 $k_B$ per molecule at 300 K. The water is modeled with the q-TIP4P/f force field \cite{Habershon2009} and the heat capacity is estimated with SC PIMD using the aforementioned DCV estimators. As shown in Figure \ref{fig:est}, both estimators converge to the same value and are in excellent agreement with the experiments. More importantly, however, the variance of the computationally expensive ${\mathcal{E}^{\square}}'$ term with the OP method is almost two orders of magnitude smaller than with the Yamamoto estimator. The OP estimator also requires one fourth of the number of force evaluations than its Yamamoto counterpart and should therefore always be preferred. 

\providecommand{\latin}[1]{#1}
\makeatletter
\providecommand{\doi}
  {\begingroup\let\do\@makeother\dospecials
  \catcode`\{=1 \catcode`\}=2 \doi@aux}
\providecommand{\doi@aux}[1]{\endgroup\texttt{#1}}
\makeatother
\providecommand*\mcitethebibliography{\thebibliography}
\csname @ifundefined\endcsname{endmcitethebibliography}
  {\let\endmcitethebibliography\endthebibliography}{}


\begin{mcitethebibliography}{104}
\providecommand*\natexlab[1]{#1}
\providecommand*\mciteSetBstSublistMode[1]{}
\providecommand*\mciteSetBstMaxWidthForm[2]{}
\providecommand*\mciteBstWouldAddEndPuncttrue
  {\def\EndOfBibitem{\unskip.}}
\providecommand*\mciteBstWouldAddEndPunctfalse
  {\let\EndOfBibitem\relax}
\providecommand*\mciteSetBstMidEndSepPunct[3]{}
\providecommand*\mciteSetBstSublistLabelBeginEnd[3]{}
\providecommand*\EndOfBibitem{}
\mciteSetBstSublistMode{f}
\mciteSetBstMaxWidthForm{subitem}{(\alph{mcitesubitemcount})}
\mciteSetBstSublistLabelBeginEnd
  {\mcitemaxwidthsubitemform\space}
  {\relax}
  {\relax}

\bibitem[Slater and Cooper(2015)Slater, and Cooper]{slater2015}
Slater,~A.~G.; Cooper,~A.~I. Function-led design of new porous materials.
  \emph{Science} \textbf{2015}, \emph{348}, aaa8075\relax
\mciteBstWouldAddEndPuncttrue
\mciteSetBstMidEndSepPunct{\mcitedefaultmidpunct}
{\mcitedefaultendpunct}{\mcitedefaultseppunct}\relax
\EndOfBibitem
\bibitem[Furukawa \latin{et~al.}(2013)Furukawa, Cordova, O'Keeffe, and
  Yaghi]{furukawa2013}
Furukawa,~H.; Cordova,~K.~E.; O'Keeffe,~M.; Yaghi,~O.~M. The chemistry and
  applications of metal-organic frameworks. \emph{Science} \textbf{2013},
  \emph{341}, 1230444\relax
\mciteBstWouldAddEndPuncttrue
\mciteSetBstMidEndSepPunct{\mcitedefaultmidpunct}
{\mcitedefaultendpunct}{\mcitedefaultseppunct}\relax
\EndOfBibitem
\bibitem[Mason \latin{et~al.}(2014)Mason, Veenstra, and Long]{mason2014}
Mason,~J.~A.; Veenstra,~M.; Long,~J.~R. Evaluating metal-organic frameworks for
  natural gas storage. \emph{Chem. Sci.} \textbf{2014}, \emph{5}, 32--51\relax
\mciteBstWouldAddEndPuncttrue
\mciteSetBstMidEndSepPunct{\mcitedefaultmidpunct}
{\mcitedefaultendpunct}{\mcitedefaultseppunct}\relax
\EndOfBibitem
\bibitem[He \latin{et~al.}(2017)He, Chen, Li, Qian, Zhou, and Chen]{he2017}
He,~Y.; Chen,~F.; Li,~B.; Qian,~G.; Zhou,~W.; Chen,~B. Porous metal-organic
  frameworks for fuel storage. \emph{Coord. Chem. Rev.} \textbf{2017},
  \emph{373}, 167--198\relax
\mciteBstWouldAddEndPuncttrue
\mciteSetBstMidEndSepPunct{\mcitedefaultmidpunct}
{\mcitedefaultendpunct}{\mcitedefaultseppunct}\relax
\EndOfBibitem
\bibitem[de~Lange \latin{et~al.}(2015)de~Lange, Verouden, Vlugt, Gascon, and
  Kapteijn]{delange2015}
de~Lange,~M.~F.; Verouden,~K. J. F.~M.; Vlugt,~T. J.~H.; Gascon,~J.;
  Kapteijn,~F. Adsorption-Driven Heat Pumps: The Potential of Metal-Organic
  Frameworks. \emph{Chem. Rev.} \textbf{2015}, \emph{115}, 12205--12250\relax
\mciteBstWouldAddEndPuncttrue
\mciteSetBstMidEndSepPunct{\mcitedefaultmidpunct}
{\mcitedefaultendpunct}{\mcitedefaultseppunct}\relax
\EndOfBibitem
\bibitem[Wang \latin{et~al.}(2018)Wang, Lee, Wahiduzzaman, Park, Muschi,
  Martineau-Corcos, Tissot, Cho, Marrot, Shepard, Maurin, Chang, and
  Serre]{wang2018}
Wang,~S.; Lee,~J.~S.; Wahiduzzaman,~M.; Park,~J.; Muschi,~M.;
  Martineau-Corcos,~C.; Tissot,~A.; Cho,~K.~H.; Marrot,~J.; Shepard,~W.;
  Maurin,~G.; Chang,~J.-S.; Serre,~C. A robust large-pore zirconium carboxylate
  metal-organic framework for energy-efficient water-sorption-driven
  refrigeration. \emph{Nat. Energy} \textbf{2018}, \emph{3}, 985--993\relax
\mciteBstWouldAddEndPuncttrue
\mciteSetBstMidEndSepPunct{\mcitedefaultmidpunct}
{\mcitedefaultendpunct}{\mcitedefaultseppunct}\relax
\EndOfBibitem
\bibitem[Sumida \latin{et~al.}(2012)Sumida, Rogow, Mason, McDonald, Bloch,
  Herm, Bae, and Long]{sumida2012}
Sumida,~K.; Rogow,~D.~L.; Mason,~J.~A.; McDonald,~T.~M.; Bloch,~E.~D.;
  Herm,~Z.~R.; Bae,~T.-H.; Long,~J.~R. Carbon Dioxide Capture in Metal-Organic
  Frameworks. \emph{Chem. Rev.} \textbf{2012}, \emph{112}, 724--781\relax
\mciteBstWouldAddEndPuncttrue
\mciteSetBstMidEndSepPunct{\mcitedefaultmidpunct}
{\mcitedefaultendpunct}{\mcitedefaultseppunct}\relax
\EndOfBibitem
\bibitem[Trickett \latin{et~al.}(2017)Trickett, Helal, Al-Maythalony, Zamani,
  Cordova, and Yaghi]{trickett2017}
Trickett,~C.~A.; Helal,~A.; Al-Maythalony,~B.~A.; Zamani,~Z.~H.;
  Cordova,~K.~E.; Yaghi,~O.~M. The chemistry of metal-organic frameworks for
  CO$_2$ capture, regeneration and conversion. \emph{Nat. Rev. Mater.}
  \textbf{2017}, \emph{2}, 17045\relax
\mciteBstWouldAddEndPuncttrue
\mciteSetBstMidEndSepPunct{\mcitedefaultmidpunct}
{\mcitedefaultendpunct}{\mcitedefaultseppunct}\relax
\EndOfBibitem
\bibitem[Simon \latin{et~al.}(2015)Simon, Kim, Gomez-Gualdron, Camp, Chung,
  Martin, Mercado, Deem, Gunter, Haranczyk, Sholl, Snurr, and Smit]{simon2015}
Simon,~C.~M.; Kim,~J.; Gomez-Gualdron,~D.~A.; Camp,~J.~S.; Chung,~Y.~G.;
  Martin,~R.~L.; Mercado,~R.; Deem,~M.~W.; Gunter,~D.; Haranczyk,~M.;
  Sholl,~D.~S.; Snurr,~R.~Q.; Smit,~B. The materials genome in action:
  identifying the performance limits for methane storage. \emph{Energy Environ.
  Sci.} \textbf{2015}, \emph{8}, 1190--1199\relax
\mciteBstWouldAddEndPuncttrue
\mciteSetBstMidEndSepPunct{\mcitedefaultmidpunct}
{\mcitedefaultendpunct}{\mcitedefaultseppunct}\relax
\EndOfBibitem
\bibitem[Mason \latin{et~al.}(2015)Mason, Oktawiec, Taylor, Hudson, Rodriguez,
  Bachman, Gonzalez, Cervellino, Guagliardi, Brown, Llewellyn, Masciocchi, and
  Long]{mason2016}
Mason,~J.~A.; Oktawiec,~J.; Taylor,~M.~K.; Hudson,~M.~R.; Rodriguez,~J.;
  Bachman,~J.~E.; Gonzalez,~M.~I.; Cervellino,~A.; Guagliardi,~A.;
  Brown,~C.~M.; Llewellyn,~P.~L.; Masciocchi,~N.; Long,~J.~R. Methane storage
  in flexible metal-organic frameworks with intrinsic thermal management.
  \emph{Nature} \textbf{2015}, \emph{527}, 357--361\relax
\mciteBstWouldAddEndPuncttrue
\mciteSetBstMidEndSepPunct{\mcitedefaultmidpunct}
{\mcitedefaultendpunct}{\mcitedefaultseppunct}\relax
\EndOfBibitem
\bibitem[Huck \latin{et~al.}(2014)Huck, Lin, Berger, Shahrak, Martin, Bhown,
  Haranczyk, Reuter, and Smit]{huck2014}
Huck,~J.~M.; Lin,~L.-C.; Berger,~A.~H.; Shahrak,~M.~N.; Martin,~R.~L.;
  Bhown,~A.~S.; Haranczyk,~M.; Reuter,~K.; Smit,~B. Evaluating different
  classes of porous materials for carbon capture. \emph{Energy Environ. Sci.}
  \textbf{2014}, \emph{7}, 4132--4146\relax
\mciteBstWouldAddEndPuncttrue
\mciteSetBstMidEndSepPunct{\mcitedefaultmidpunct}
{\mcitedefaultendpunct}{\mcitedefaultseppunct}\relax
\EndOfBibitem
\bibitem[Mu and Walton(2011)Mu, and Walton]{mu2011}
Mu,~B.; Walton,~K.~S. Thermal Analysis and Heat Capacity Study of Metal-Organic
  Frameworks. \emph{J. Phys. Chem. C} \textbf{2011}, \emph{115},
  22748--22754\relax
\mciteBstWouldAddEndPuncttrue
\mciteSetBstMidEndSepPunct{\mcitedefaultmidpunct}
{\mcitedefaultendpunct}{\mcitedefaultseppunct}\relax
\EndOfBibitem
\bibitem[Kloutse \latin{et~al.}(2015)Kloutse, Zacharia, Cossement, and
  Chahine]{kloutse2015}
Kloutse,~F.~A.; Zacharia,~R.; Cossement,~D.; Chahine,~R. Specific heat
  capacities of {MOF}-5, {C}u-{BTC}, {F}e-{BTC}, {MOF}-177 and {MIL}-53({A}l)
  over wide temperature ranges: Measurements and application of empirical group
  contribution method. \emph{Microporous Mesoporous Mater.} \textbf{2015},
  \emph{217}, 1--5\relax
\mciteBstWouldAddEndPuncttrue
\mciteSetBstMidEndSepPunct{\mcitedefaultmidpunct}
{\mcitedefaultendpunct}{\mcitedefaultseppunct}\relax
\EndOfBibitem
\bibitem[Babaei and Wilmer(2016)Babaei, and Wilmer]{babaei2016}
Babaei,~H.; Wilmer,~C.~E. Mechanisms of Heat Transfer in Porous Crystals
  Containing Adsorbed Gases: Applications to Metal-Organic Frameworks.
  \emph{Phys. Rev. Lett.} \textbf{2016}, \emph{116}, 025902\relax
\mciteBstWouldAddEndPuncttrue
\mciteSetBstMidEndSepPunct{\mcitedefaultmidpunct}
{\mcitedefaultendpunct}{\mcitedefaultseppunct}\relax
\EndOfBibitem
\bibitem[Balestra \latin{et~al.}(2016)Balestra, Bueno-Perez, Hamad, Dubbeldam,
  Ruiz-Salvador, and Calero]{balestra2016}
Balestra,~S. R.~G.; Bueno-Perez,~R.; Hamad,~S.; Dubbeldam,~D.;
  Ruiz-Salvador,~A.~R.; Calero,~S. Controlling Thermal Expansion: A
  Metal-Organic Frameworks Route. \emph{Chem. Mater.} \textbf{2016}, \emph{28},
  8296--8304\relax
\mciteBstWouldAddEndPuncttrue
\mciteSetBstMidEndSepPunct{\mcitedefaultmidpunct}
{\mcitedefaultendpunct}{\mcitedefaultseppunct}\relax
\EndOfBibitem
\bibitem[Auckett \latin{et~al.}(2018)Auckett, Barkhordarian, Ogilvie, Duyker,
  Chevreau, Peterson, and Kepert]{auckett2018}
Auckett,~J.~E.; Barkhordarian,~A.~A.; Ogilvie,~S.~H.; Duyker,~S.~G.;
  Chevreau,~H.; Peterson,~V.~K.; Kepert,~C.~J. Continuous negative-to-positive
  tuning of thermal expansion achieved by controlled gas sorption in porous
  coordination frameworks. \emph{Nat. Commun.} \textbf{2018}, \emph{9},
  4873\relax
\mciteBstWouldAddEndPuncttrue
\mciteSetBstMidEndSepPunct{\mcitedefaultmidpunct}
{\mcitedefaultendpunct}{\mcitedefaultseppunct}\relax
\EndOfBibitem
\bibitem[Rogge \latin{et~al.}(2015)Rogge, Vanduyfhuys, Ghysels, Waroquier,
  Verstraelen, Maurin, and Van~Speybroeck]{rogge2015}
Rogge,~S. M.~J.; Vanduyfhuys,~L.; Ghysels,~A.; Waroquier,~M.; Verstraelen,~T.;
  Maurin,~G.; Van~Speybroeck,~V. A Comparison of Barostats for the Mechanical
  Characterization of Metal-Organic Frameworks. \emph{J. Chem. Theory Comput.}
  \textbf{2015}, \emph{11}, 5583--5597\relax
\mciteBstWouldAddEndPuncttrue
\mciteSetBstMidEndSepPunct{\mcitedefaultmidpunct}
{\mcitedefaultendpunct}{\mcitedefaultseppunct}\relax
\EndOfBibitem
\bibitem[Vanduyfhuys \latin{et~al.}(2018)Vanduyfhuys, Rogge, Wieme,
  Vandenbrande, Maurin, Waroquier, and Van~Speybroeck]{vanduyfhuys2018b}
Vanduyfhuys,~L.; Rogge,~S. M.~J.; Wieme,~J.; Vandenbrande,~S.; Maurin,~G.;
  Waroquier,~M.; Van~Speybroeck,~V. Thermodynamic insight into
  stimuli-responsive behaviour of soft porous crystals. \emph{Nat. Commun.}
  \textbf{2018}, \emph{9}, 204\relax
\mciteBstWouldAddEndPuncttrue
\mciteSetBstMidEndSepPunct{\mcitedefaultmidpunct}
{\mcitedefaultendpunct}{\mcitedefaultseppunct}\relax
\EndOfBibitem
\bibitem[Wieme \latin{et~al.}(2018)Wieme, Lejaeghere, Kresse, and
  Van~Speybroeck]{wieme2018}
Wieme,~J.; Lejaeghere,~K.; Kresse,~G.; Van~Speybroeck,~V. Tuning the balance
  between dispersion and entropy to design temperature-responsive flexible
  metal-organic frameworks. \emph{Nat. Commun.} \textbf{2018}, \emph{9},
  4899\relax
\mciteBstWouldAddEndPuncttrue
\mciteSetBstMidEndSepPunct{\mcitedefaultmidpunct}
{\mcitedefaultendpunct}{\mcitedefaultseppunct}\relax
\EndOfBibitem
\bibitem[Paesani(2012)]{paesani2012}
Paesani,~F. Water in metal-organic frameworks: structure and diffusion of
  H$_2$O in MIL-53(Cr) from quantum simulations. \emph{Mol. Simul.}
  \textbf{2012}, \emph{38}, 631--641\relax
\mciteBstWouldAddEndPuncttrue
\mciteSetBstMidEndSepPunct{\mcitedefaultmidpunct}
{\mcitedefaultendpunct}{\mcitedefaultseppunct}\relax
\EndOfBibitem
\bibitem[Borges \latin{et~al.}(2017)Borges, Semino, Devautour-Vinot, Jobic,
  Paesani, and Maurin]{borges2017}
Borges,~D.~D.; Semino,~R.; Devautour-Vinot,~S.; Jobic,~H.; Paesani,~F.;
  Maurin,~G. Computational Exploration of the Water Concentration Dependence of
  the Proton Transport in the Porous UiO-66(Zr)-(CO$_2$H)$_2$ Metal-Organic
  Framework. \emph{Chem. Mater.} \textbf{2017}, \emph{29}, 1569--1576\relax
\mciteBstWouldAddEndPuncttrue
\mciteSetBstMidEndSepPunct{\mcitedefaultmidpunct}
{\mcitedefaultendpunct}{\mcitedefaultseppunct}\relax
\EndOfBibitem
\bibitem[Parrinello and Rahman(1984)Parrinello, and Rahman]{parr-rahm84jcp}
Parrinello,~M.; Rahman,~A. {Study of an F center in molten KCl}. \emph{J. Chem.
  Phys.} \textbf{1984}, \emph{80}, 860\relax
\mciteBstWouldAddEndPuncttrue
\mciteSetBstMidEndSepPunct{\mcitedefaultmidpunct}
{\mcitedefaultendpunct}{\mcitedefaultseppunct}\relax
\EndOfBibitem
\bibitem[Chandler and Wolynes(1981)Chandler, and Wolynes]{chan-woly81jcp}
Chandler,~D.; Wolynes,~P.~G. {Exploiting the isomorphism between quantum theory
  and classical statistical mechanics of polyatomic fluids}. \emph{J. Chem.
  Phys.} \textbf{1981}, \emph{74}, 4078--4095\relax
\mciteBstWouldAddEndPuncttrue
\mciteSetBstMidEndSepPunct{\mcitedefaultmidpunct}
{\mcitedefaultendpunct}{\mcitedefaultseppunct}\relax
\EndOfBibitem
\bibitem[Jang \latin{et~al.}(2001)Jang, Jang, and Voth]{jang-voth01jcp}
Jang,~S.; Jang,~S.; Voth,~G.~A. {Applications of higher order composite
  factorization schemes in imaginary time path integral simulations}. \emph{J.
  Chem. Phys.} \textbf{2001}, \emph{115}, 7832--7842\relax
\mciteBstWouldAddEndPuncttrue
\mciteSetBstMidEndSepPunct{\mcitedefaultmidpunct}
{\mcitedefaultendpunct}{\mcitedefaultseppunct}\relax
\EndOfBibitem
\bibitem[Ceriotti and Manolopoulos(2012)Ceriotti, and Manolopoulos]{ceri12prl}
Ceriotti,~M.; Manolopoulos,~D.~E. Efficient First-Principles Calculation of the
  Quantum Kinetic Energy and Momentum Distribution of Nuclei. \emph{Phys. Rev.
  Lett.} \textbf{2012}, \emph{109}, 100604\relax
\mciteBstWouldAddEndPuncttrue
\mciteSetBstMidEndSepPunct{\mcitedefaultmidpunct}
{\mcitedefaultendpunct}{\mcitedefaultseppunct}\relax
\EndOfBibitem
\bibitem[Kapil \latin{et~al.}(2016)Kapil, Behler, and Ceriotti]{kapi+16jcp2}
Kapil,~V.; Behler,~J.; Ceriotti,~M. {High order path integrals made easy}.
  \emph{J. Chem. Phys.} \textbf{2016}, \emph{145}, 234103\relax
\mciteBstWouldAddEndPuncttrue
\mciteSetBstMidEndSepPunct{\mcitedefaultmidpunct}
{\mcitedefaultendpunct}{\mcitedefaultseppunct}\relax
\EndOfBibitem
\bibitem[Markland and Manolopoulos(2008)Markland, and
  Manolopoulos]{mark-mano08jcp}
Markland,~T.~E.; Manolopoulos,~D.~E. {An efficient ring polymer contraction
  scheme for imaginary time path integral simulations.} \emph{J. Chem. Phys.}
  \textbf{2008}, \emph{129}, 024105\relax
\mciteBstWouldAddEndPuncttrue
\mciteSetBstMidEndSepPunct{\mcitedefaultmidpunct}
{\mcitedefaultendpunct}{\mcitedefaultseppunct}\relax
\EndOfBibitem
\bibitem[Kapil \latin{et~al.}(2016)Kapil, VandeVondele, and
  Ceriotti]{kapi+16jcp}
Kapil,~V.; VandeVondele,~J.; Ceriotti,~M. {Accurate molecular dynamics and
  nuclear quantum effects at low cost by multiple steps in real and imaginary
  time: Using density functional theory to accelerate wavefunction methods}.
  \emph{J. Chem. Phys.} \textbf{2016}, \emph{144}, 054111\relax
\mciteBstWouldAddEndPuncttrue
\mciteSetBstMidEndSepPunct{\mcitedefaultmidpunct}
{\mcitedefaultendpunct}{\mcitedefaultseppunct}\relax
\EndOfBibitem
\bibitem[Poltavsky and Tkatchenko(2016)Poltavsky, and
  Tkatchenko]{polt-tkat16cs}
Poltavsky,~I.; Tkatchenko,~A. {Modeling quantum nuclei with perturbed path
  integral molecular dynamics}. \emph{Chem. Sci.} \textbf{2016}, \emph{7},
  1368--1372\relax
\mciteBstWouldAddEndPuncttrue
\mciteSetBstMidEndSepPunct{\mcitedefaultmidpunct}
{\mcitedefaultendpunct}{\mcitedefaultseppunct}\relax
\EndOfBibitem
\bibitem[Markland and Ceriotti(2018)Markland, and Ceriotti]{mark-ceri18nrc}
Markland,~T.~E.; Ceriotti,~M. {Nuclear quantum effects enter the mainstream}.
  \emph{Nat. Rev. Chem.} \textbf{2018}, \emph{2}, 0109\relax
\mciteBstWouldAddEndPuncttrue
\mciteSetBstMidEndSepPunct{\mcitedefaultmidpunct}
{\mcitedefaultendpunct}{\mcitedefaultseppunct}\relax
\EndOfBibitem
\bibitem[Rogge \latin{et~al.}(2018)Rogge, Caroes, Demuynck, Waroquier,
  Speybroeck, and Ghysels]{Rogge2018}
Rogge,~S. M.~J.; Caroes,~S.; Demuynck,~R.; Waroquier,~M.; Speybroeck,~V.~V.;
  Ghysels,~A. The Importance of Cell Shape Sampling To Accurately Predict
  Flexibility in Metal{\textendash}Organic Frameworks. \emph{J. Chem. Theory
  Comput.} \textbf{2018}, \emph{14}, 1186--1197\relax
\mciteBstWouldAddEndPuncttrue
\mciteSetBstMidEndSepPunct{\mcitedefaultmidpunct}
{\mcitedefaultendpunct}{\mcitedefaultseppunct}\relax
\EndOfBibitem
\bibitem[Martyna \latin{et~al.}(1999)Martyna, Hughes, and
  Tuckerman]{mart+99jcp}
Martyna,~G.~J.; Hughes,~A.; Tuckerman,~M.~E. {Molecular dynamics algorithms for
  path integrals at constant pressure}. \emph{J. Chem. Phys.} \textbf{1999},
  \emph{110}, 3275\relax
\mciteBstWouldAddEndPuncttrue
\mciteSetBstMidEndSepPunct{\mcitedefaultmidpunct}
{\mcitedefaultendpunct}{\mcitedefaultseppunct}\relax
\EndOfBibitem
\bibitem[Ceriotti \latin{et~al.}(2014)Ceriotti, More, and Manolopoulos]{ipi1}
Ceriotti,~M.; More,~J.; Manolopoulos,~D.~E. i-PI: A Python interface for ab
  initio path integral molecular dynamics simulations. \emph{Comp. Phys.
  Commun.} \textbf{2014}, \emph{185}, 1019--1026\relax
\mciteBstWouldAddEndPuncttrue
\mciteSetBstMidEndSepPunct{\mcitedefaultmidpunct}
{\mcitedefaultendpunct}{\mcitedefaultseppunct}\relax
\EndOfBibitem
\bibitem[Vanduyfhuys \latin{et~al.}(2015)Vanduyfhuys, Vandenbrande,
  Verstraelen, Schmid, Waroquier, and Van~Speybroeck]{vanduyfhuys2015}
Vanduyfhuys,~L.; Vandenbrande,~S.; Verstraelen,~T.; Schmid,~R.; Waroquier,~M.;
  Van~Speybroeck,~V. Quick{FF}: A program for a quick and easy derivation of
  force fields for metal-organic frameworks from ab initio input. \emph{J.
  Comput. Chem.} \textbf{2015}, \emph{36}, 1015--1027\relax
\mciteBstWouldAddEndPuncttrue
\mciteSetBstMidEndSepPunct{\mcitedefaultmidpunct}
{\mcitedefaultendpunct}{\mcitedefaultseppunct}\relax
\EndOfBibitem
\bibitem[Vanduyfhuys \latin{et~al.}(2018)Vanduyfhuys, Vandenbrande, Wieme,
  Waroquier, Verstraelen, and Van~Speybroeck]{vanduyfhuys2018}
Vanduyfhuys,~L.; Vandenbrande,~S.; Wieme,~J.; Waroquier,~M.; Verstraelen,~T.;
  Van~Speybroeck,~V. Extension of the Quick{FF} force field protocol for an
  improved accuracy of structural, vibrational, mechanical and thermal
  properties of metal-organic frameworks. \emph{J. Comput. Chem.}
  \textbf{2018}, \emph{39}, 999--1011\relax
\mciteBstWouldAddEndPuncttrue
\mciteSetBstMidEndSepPunct{\mcitedefaultmidpunct}
{\mcitedefaultendpunct}{\mcitedefaultseppunct}\relax
\EndOfBibitem
\bibitem[Li \latin{et~al.}(1999)Li, Eddaoudi, O'Keeffe, and Yaghi]{li1999}
Li,~H.; Eddaoudi,~M.; O'Keeffe,~M.; Yaghi,~O.~M. Design and synthesis of an
  exceptionally stable and highly porous metal-organic framework. \emph{Nature}
  \textbf{1999}, \emph{402}, 276--279\relax
\mciteBstWouldAddEndPuncttrue
\mciteSetBstMidEndSepPunct{\mcitedefaultmidpunct}
{\mcitedefaultendpunct}{\mcitedefaultseppunct}\relax
\EndOfBibitem
\bibitem[Eddaoudi \latin{et~al.}(2002)Eddaoudi, Kim, Rosi, Vodak, Wachter,
  O'Keeffe, and Yaghi]{eddaoudi2002}
Eddaoudi,~M.; Kim,~J.; Rosi,~N.; Vodak,~D.; Wachter,~J.; O'Keeffe,~M.;
  Yaghi,~O.~M. Systematic Design of Pore Size and Functionality in Isoreticular
  MOFs and Their Application in Methane Storage. \emph{Science} \textbf{2002},
  \emph{295}, 469--472\relax
\mciteBstWouldAddEndPuncttrue
\mciteSetBstMidEndSepPunct{\mcitedefaultmidpunct}
{\mcitedefaultendpunct}{\mcitedefaultseppunct}\relax
\EndOfBibitem
\bibitem[Zhou \latin{et~al.}(2007)Zhou, Wu, Hartman, and Yildirim]{zhou2007}
Zhou,~W.; Wu,~H.; Hartman,~M.~R.; Yildirim,~T. Hydrogen and Methane Adsorption
  in Metal-Organic Frameworks: A High-Pressure Volumetric Study. \emph{J. Phys.
  Chem. C} \textbf{2007}, \emph{111}, 16131--16137\relax
\mciteBstWouldAddEndPuncttrue
\mciteSetBstMidEndSepPunct{\mcitedefaultmidpunct}
{\mcitedefaultendpunct}{\mcitedefaultseppunct}\relax
\EndOfBibitem
\bibitem[Martyna \latin{et~al.}(1994)Martyna, Tobias, and Klein]{mart+94jcp}
Martyna,~G.~J.; Tobias,~D.~J.; Klein,~M.~L. {Constant pressure molecular
  dynamics algorithms}. \emph{J. Chem. Phys.} \textbf{1994}, \emph{101},
  4177\relax
\mciteBstWouldAddEndPuncttrue
\mciteSetBstMidEndSepPunct{\mcitedefaultmidpunct}
{\mcitedefaultendpunct}{\mcitedefaultseppunct}\relax
\EndOfBibitem
\bibitem[Rahman(1964)]{Rahman1964}
Rahman,~A. Correlations in the Motion of Atoms in Liquid Argon. \emph{Physical
  Review} \textbf{1964}, \emph{136}, A405--A411\relax
\mciteBstWouldAddEndPuncttrue
\mciteSetBstMidEndSepPunct{\mcitedefaultmidpunct}
{\mcitedefaultendpunct}{\mcitedefaultseppunct}\relax
\EndOfBibitem
\bibitem[Uhl \latin{et~al.}(2016)Uhl, Marx, and Ceriotti]{uhl+16jcp}
Uhl,~F.; Marx,~D.; Ceriotti,~M. {Accelerated path integral methods for
  atomistic simulations at ultra-low temperatures}. \emph{J. Chem. Phys.}
  \textbf{2016}, \emph{145}, 054101\relax
\mciteBstWouldAddEndPuncttrue
\mciteSetBstMidEndSepPunct{\mcitedefaultmidpunct}
{\mcitedefaultendpunct}{\mcitedefaultseppunct}\relax
\EndOfBibitem
\bibitem[Shiga and Shinoda(2005)Shiga, and Shinoda]{shiga2005}
Shiga,~M.; Shinoda,~W. Calculation of heat capacities of light and heavy water
  by path-integral molecular dynamics. \emph{J. Chem. Phys.} \textbf{2005},
  \emph{123}, 134502\relax
\mciteBstWouldAddEndPuncttrue
\mciteSetBstMidEndSepPunct{\mcitedefaultmidpunct}
{\mcitedefaultendpunct}{\mcitedefaultseppunct}\relax
\EndOfBibitem
\bibitem[Takahashi and Imada(1984)Takahashi, and Imada]{taka-imad84jpsj}
Takahashi,~M.; Imada,~M. {Monte Carlo calculation of quantum systems. II.
  Higher order correction}. \emph{J. Phys. Soc. Jap.} \textbf{1984}, \emph{53},
  3765--3769\relax
\mciteBstWouldAddEndPuncttrue
\mciteSetBstMidEndSepPunct{\mcitedefaultmidpunct}
{\mcitedefaultendpunct}{\mcitedefaultseppunct}\relax
\EndOfBibitem
\bibitem[Chin(1997)]{chin97pla}
Chin,~S.~A. {Symplectic integrators from composite operator factorizations}.
  \emph{Phys. Lett. A} \textbf{1997}, \emph{226}, 344--348\relax
\mciteBstWouldAddEndPuncttrue
\mciteSetBstMidEndSepPunct{\mcitedefaultmidpunct}
{\mcitedefaultendpunct}{\mcitedefaultseppunct}\relax
\EndOfBibitem
\bibitem[alp()]{alpha_parameter}
The standard fourth order Suzuki-Chin Splitting contains a $alpha$ parameter
  which can be used to share the high order term between odd and even beads. In
  this case we have set it to zero so that the high order term is evaluated
  only on the even beads reducing the computational cost by a factor of
  two.\relax
\mciteBstWouldAddEndPunctfalse
\mciteSetBstMidEndSepPunct{\mcitedefaultmidpunct}
{}{\mcitedefaultseppunct}\relax
\EndOfBibitem
\bibitem[Suzuki(1995)]{suzu95pla}
Suzuki,~M. {Hybrid exponential product formulas for unbounded operators with
  possible applications to Monte Carlo simulations}. \emph{Phys. Lett. A}
  \textbf{1995}, \emph{201}, 425--428\relax
\mciteBstWouldAddEndPuncttrue
\mciteSetBstMidEndSepPunct{\mcitedefaultmidpunct}
{\mcitedefaultendpunct}{\mcitedefaultseppunct}\relax
\EndOfBibitem
\bibitem[Brain(2011)]{brainthesis}
Brain,~G. Higher Order Propagators in Path Integral Molecular Dynamics. Ph.D.\
  thesis, Part II Chemistry Thesis, Oxford University, 2011\relax
\mciteBstWouldAddEndPuncttrue
\mciteSetBstMidEndSepPunct{\mcitedefaultmidpunct}
{\mcitedefaultendpunct}{\mcitedefaultseppunct}\relax
\EndOfBibitem
\bibitem[Jang \latin{et~al.}(2014)Jang, Sinitskiy, and Voth]{jang+14jcp}
Jang,~S.; Sinitskiy,~A.~V.; Voth,~G.~a. {Can the ring polymer molecular
  dynamics method be interpreted as real time quantum dynamics?} \emph{J. Chem.
  Phys.} \textbf{2014}, \emph{140}, 154103\relax
\mciteBstWouldAddEndPuncttrue
\mciteSetBstMidEndSepPunct{\mcitedefaultmidpunct}
{\mcitedefaultendpunct}{\mcitedefaultseppunct}\relax
\EndOfBibitem
\bibitem[P{\'{e}}rez and Tuckerman(2011)P{\'{e}}rez, and
  Tuckerman]{pere-tuck11jcp}
P{\'{e}}rez,~A.; Tuckerman,~M.~E. {Improving the convergence of closed and open
  path integral molecular dynamics via higher order Trotter factorization
  schemes.} \emph{J. Chem. Phys.} \textbf{2011}, \emph{135}, 064104\relax
\mciteBstWouldAddEndPuncttrue
\mciteSetBstMidEndSepPunct{\mcitedefaultmidpunct}
{\mcitedefaultendpunct}{\mcitedefaultseppunct}\relax
\EndOfBibitem
\bibitem[Yamamoto(2005)]{yama05jcp}
Yamamoto,~T.~M. {Path-integral virial estimator based on the scaling of
  fluctuation coordinates: Application to quantum clusters with fourth-order
  propagators}. \emph{J. Chem. Phys.} \textbf{2005}, \emph{123}, 104101\relax
\mciteBstWouldAddEndPuncttrue
\mciteSetBstMidEndSepPunct{\mcitedefaultmidpunct}
{\mcitedefaultendpunct}{\mcitedefaultseppunct}\relax
\EndOfBibitem
\bibitem[Ceriotti \latin{et~al.}(2011)Ceriotti, Brain, Riordan, and
  Manolopoulos]{ceri+12prsa}
Ceriotti,~M.; Brain,~G. A.~R.; Riordan,~O.; Manolopoulos,~D.~E. {The
  inefficiency of re-weighted sampling and the curse of system size in high
  order path integration}. \emph{Proc. Royal Soc. A} \textbf{2011}, \emph{468},
  2--17\relax
\mciteBstWouldAddEndPuncttrue
\mciteSetBstMidEndSepPunct{\mcitedefaultmidpunct}
{\mcitedefaultendpunct}{\mcitedefaultseppunct}\relax
\EndOfBibitem
\bibitem[Kapil \latin{et~al.}(2018)Kapil, Rossi, Marsalek, Petraglia, Litman,
  Spura, Cheng, Cuzzocrea, Mei{\ss}ner, Wilkins, Juda, Bienvenue, Fang,
  Kessler, Poltavsky, Vandenbrande, Wieme, Corminboeuf, K\"{u}hne,
  Manolopoulos, Markland, Richardson, Tkatchenko, Tribello, Speybroeck, and
  Ceriotti]{ipi2}
Kapil,~V. \latin{et~al.}  i-{PI} 2.0: A universal force engine for advanced
  molecular simulations. \emph{Comput. Phys. Commun.} \textbf{2018}, \emph{in
  press}\relax
\mciteBstWouldAddEndPuncttrue
\mciteSetBstMidEndSepPunct{\mcitedefaultmidpunct}
{\mcitedefaultendpunct}{\mcitedefaultseppunct}\relax
\EndOfBibitem
\bibitem[Tuckerman \latin{et~al.}(1992)Tuckerman, Berne, and
  Martyna]{tuck+92jcp}
Tuckerman,~M.; Berne,~B.~J.; Martyna,~G.~J. {Reversible multiple time scale
  molecular dynamics}. \emph{J. Chem. Phys.} \textbf{1992}, \emph{97},
  1990\relax
\mciteBstWouldAddEndPuncttrue
\mciteSetBstMidEndSepPunct{\mcitedefaultmidpunct}
{\mcitedefaultendpunct}{\mcitedefaultseppunct}\relax
\EndOfBibitem
\bibitem[Leimkuhler and Matthews(2013)Leimkuhler, and Matthews]{leim+13jcp}
Leimkuhler,~B.; Matthews,~C. {Robust and efficient configurational molecular
  sampling via Langevin dynamics}. \emph{J. Chem. Phys.} \textbf{2013},
  \emph{138}\relax
\mciteBstWouldAddEndPuncttrue
\mciteSetBstMidEndSepPunct{\mcitedefaultmidpunct}
{\mcitedefaultendpunct}{\mcitedefaultseppunct}\relax
\EndOfBibitem
\bibitem[Kapil \latin{et~al.}(2018)Kapil, Cuzzocrea, and Ceriotti]{kapi+18jpcb}
Kapil,~V.; Cuzzocrea,~A.; Ceriotti,~M. {Anisotropy of the Proton Momentum
  Distribution in Water}. \emph{J. Phys. Chem. B} \textbf{2018}, \emph{122},
  6048--6054\relax
\mciteBstWouldAddEndPuncttrue
\mciteSetBstMidEndSepPunct{\mcitedefaultmidpunct}
{\mcitedefaultendpunct}{\mcitedefaultseppunct}\relax
\EndOfBibitem
\bibitem[Poltavsky \latin{et~al.}(2018)Poltavsky, Jr., and
  Tkatchenko]{Poltavsky2018}
Poltavsky,~I.; Jr.,~R. A.~D.; Tkatchenko,~A. Perturbed path integrals in
  imaginary time: Efficiently modeling nuclear quantum effects in molecules and
  materials. \emph{J. Chem. Phys.} \textbf{2018}, \emph{148}, 102325\relax
\mciteBstWouldAddEndPuncttrue
\mciteSetBstMidEndSepPunct{\mcitedefaultmidpunct}
{\mcitedefaultendpunct}{\mcitedefaultseppunct}\relax
\EndOfBibitem
\bibitem[Yamamoto(2005)]{yamamoto2005}
Yamamoto,~T.~M. Path-integral virial estimator based on the scaling of
  fluctuation coordinates: Application to quantum clusters with fourth-order
  propagators. \emph{J. Chem. Phys.} \textbf{2005}, \emph{123}, 104101\relax
\mciteBstWouldAddEndPuncttrue
\mciteSetBstMidEndSepPunct{\mcitedefaultmidpunct}
{\mcitedefaultendpunct}{\mcitedefaultseppunct}\relax
\EndOfBibitem
\bibitem[Tafipolsky \latin{et~al.}(2007)Tafipolsky, Amirjalayer, and
  Schmid]{tafipolsky2007}
Tafipolsky,~M.; Amirjalayer,~S.; Schmid,~R. Ab initio parametrized {MM}3 force
  field for the metal-organic framework {MOF}-5. \emph{J. Comput. Chem.}
  \textbf{2007}, \emph{28}, 1169--1176\relax
\mciteBstWouldAddEndPuncttrue
\mciteSetBstMidEndSepPunct{\mcitedefaultmidpunct}
{\mcitedefaultendpunct}{\mcitedefaultseppunct}\relax
\EndOfBibitem
\bibitem[Frisch \latin{et~al.}(2016)Frisch, Trucks, Schlegel, Scuseria, Robb,
  Cheeseman, Scalmani, Barone, Petersson, Nakatsuji, Li, Caricato, Marenich,
  Bloino, Janesko, Gomperts, Mennucci, Hratchian, Ortiz, Izmaylov, Sonnenberg,
  Williams-Young, Ding, Lipparini, Egidi, Goings, Peng, Petrone, Henderson,
  Ranasinghe, Zakrzewski, Gao, Rega, Zheng, Liang, Hada, Ehara, Toyota, Fukuda,
  Hasegawa, Ishida, Nakajima, Honda, Kitao, Nakai, Vreven, Throssell,
  Montgomery, Peralta, Ogliaro, Bearpark, Heyd, Brothers, Kudin, Staroverov,
  Keith, Kobayashi, Normand, Raghavachari, Rendell, Burant, Iyengar, Tomasi,
  Cossi, Millam, Klene, Adamo, Cammi, Ochterski, Martin, Morokuma, Farkas,
  Foresman, and Fox]{gaussian16}
Frisch,~M.~J. \latin{et~al.}  Gaussian16 {R}evision {B}.01. 2016; Gaussian Inc.
  Wallingford CT\relax
\mciteBstWouldAddEndPuncttrue
\mciteSetBstMidEndSepPunct{\mcitedefaultmidpunct}
{\mcitedefaultendpunct}{\mcitedefaultseppunct}\relax
\EndOfBibitem
\bibitem[Becke(1988)]{b3lypa}
Becke,~A.~D. Density-Functional Exchange-Energy Approximation with Correct
  Asymptotic Behavior. \emph{Phys. Rev. A} \textbf{1988}, \emph{38},
  3098--3100\relax
\mciteBstWouldAddEndPuncttrue
\mciteSetBstMidEndSepPunct{\mcitedefaultmidpunct}
{\mcitedefaultendpunct}{\mcitedefaultseppunct}\relax
\EndOfBibitem
\bibitem[Becke(1993)]{b3lypb}
Becke,~A.~D. Density-Functional Thermochemistry. III. The Role of Exact
  Exchange. \emph{J. Chem. Phys.} \textbf{1993}, \emph{98}, 5648--5652\relax
\mciteBstWouldAddEndPuncttrue
\mciteSetBstMidEndSepPunct{\mcitedefaultmidpunct}
{\mcitedefaultendpunct}{\mcitedefaultseppunct}\relax
\EndOfBibitem
\bibitem[Lee \latin{et~al.}(1988)Lee, Yang, and Parr]{b3lypc}
Lee,~C.; Yang,~W.; Parr,~R.~G. Development of the Colle-Salvetti
  Correlation-Energy Formula into a Functional of the Electron Density.
  \emph{Phys. Rev. B} \textbf{1988}, \emph{37}, 785--789\relax
\mciteBstWouldAddEndPuncttrue
\mciteSetBstMidEndSepPunct{\mcitedefaultmidpunct}
{\mcitedefaultendpunct}{\mcitedefaultseppunct}\relax
\EndOfBibitem
\bibitem[Krishnan \latin{et~al.}(1980)Krishnan, Binkley, Seeger, and
  Pople]{Poplebasisset}
Krishnan,~R.; Binkley,~J.~S.; Seeger,~R.; Pople,~J.~A. Self-Consistent
  Molecular Orbital Methods. 20. Basis Set for Correlated wave-Functions.
  \emph{J. Chem. Phys.} \textbf{1980}, \emph{72}, 650--654\relax
\mciteBstWouldAddEndPuncttrue
\mciteSetBstMidEndSepPunct{\mcitedefaultmidpunct}
{\mcitedefaultendpunct}{\mcitedefaultseppunct}\relax
\EndOfBibitem
\bibitem[Hay and Wadt(1985)Hay, and Wadt]{Lanl}
Hay,~P.~J.; Wadt,~W.~R. Ab initio effective core potentials for molecular
  calculations. Potentials for the transition metal atoms Sc to Hg. \emph{J.
  Chem. Phys.} \textbf{1985}, \emph{82}, 270\relax
\mciteBstWouldAddEndPuncttrue
\mciteSetBstMidEndSepPunct{\mcitedefaultmidpunct}
{\mcitedefaultendpunct}{\mcitedefaultseppunct}\relax
\EndOfBibitem
\bibitem[Verstraelen \latin{et~al.}(2016)Verstraelen, Vandenbrande,
  Heidar-Zadeh, Vanduyfhuys, Van~Speybroeck, Waroquier, and
  Ayers]{verstraelen2016}
Verstraelen,~T.; Vandenbrande,~S.; Heidar-Zadeh,~F.; Vanduyfhuys,~L.;
  Van~Speybroeck,~V.; Waroquier,~M.; Ayers,~P.~W. Minimal Basis Iterative
  Stockholder: Atoms in Molecules for Force-Field Development. \emph{J. Chem.
  Theory Comput.} \textbf{2016}, \emph{12}, 3894--3912\relax
\mciteBstWouldAddEndPuncttrue
\mciteSetBstMidEndSepPunct{\mcitedefaultmidpunct}
{\mcitedefaultendpunct}{\mcitedefaultseppunct}\relax
\EndOfBibitem
\bibitem[Perdew \latin{et~al.}(1996)Perdew, Burke, and Ernzerhof]{perdew1996}
Perdew,~J.~P.; Burke,~K.; Ernzerhof,~M. Generalized Gradient Approximation Made
  Simple. \emph{Phys. Rev. Lett.} \textbf{1996}, \emph{77}, 3865\relax
\mciteBstWouldAddEndPuncttrue
\mciteSetBstMidEndSepPunct{\mcitedefaultmidpunct}
{\mcitedefaultendpunct}{\mcitedefaultseppunct}\relax
\EndOfBibitem
\bibitem[Mortensen \latin{et~al.}(2005)Mortensen, Hansen, and Jacobsen]{gpaw1}
Mortensen,~J.~J.; Hansen,~L.~B.; Jacobsen,~K.~W. Real-space grid implementation
  of the projector augmented wave method. \emph{Phys. Rev. B} \textbf{2005},
  \emph{71}, 035109\relax
\mciteBstWouldAddEndPuncttrue
\mciteSetBstMidEndSepPunct{\mcitedefaultmidpunct}
{\mcitedefaultendpunct}{\mcitedefaultseppunct}\relax
\EndOfBibitem
\bibitem[Enkovaara \latin{et~al.}(2010)Enkovaara, Rostgaard, Mortensen, Chen,
  Du\l{}ak, Ferrighi, Gavnholt, Glinsvad, Haikola, Hansen, Kristoffersen,
  Kuisma, Larsen, Lehtovaara, Ljungberg, Lopez-Acevedo, Moses, Ojanen, Olsen,
  Petzold, Romero, Stausholm-M{\o}ller, Strange, Tritsaris, Vanin, Walter,
  Hammer, H\"{a}kkinen, Madsen, Nieminen, N{\o}rskov, Puska, Rantala,
  Schi{\o}tz, Thygesen, and Jacobsen]{gpaw2}
Enkovaara,~J. \latin{et~al.}  Electronic structure calculations with GPAW: a
  real-space implementation of the projector augmented-wave method. \emph{J.
  Phys.: Condens. Matter} \textbf{2010}, \emph{22}, 253202\relax
\mciteBstWouldAddEndPuncttrue
\mciteSetBstMidEndSepPunct{\mcitedefaultmidpunct}
{\mcitedefaultendpunct}{\mcitedefaultseppunct}\relax
\EndOfBibitem
\bibitem[Verstraelen \latin{et~al.}(2017)Verstraelen, Tecmer, Heidar-Zadeh,
  Gonz\'{a}lez-Espinoza, Kim, Boguslawski, Fias, Vandenbrande, Berrocal, and
  Ayers]{horton}
Verstraelen,~T.; Tecmer,~P.; Heidar-Zadeh,~F.; Gonz\'{a}lez-Espinoza,~C.~E.;
  Kim,~T.~D.; Boguslawski,~K.; Fias,~S.; Vandenbrande,~S.; Berrocal,~D.;
  Ayers,~P.~W. Horton 2.1.0. 2017;
  \url{https://theochem.github.io/horton/2.1.0/index.html}\relax
\mciteBstWouldAddEndPuncttrue
\mciteSetBstMidEndSepPunct{\mcitedefaultmidpunct}
{\mcitedefaultendpunct}{\mcitedefaultseppunct}\relax
\EndOfBibitem
\bibitem[Lii and Allinger(1989)Lii, and Allinger]{lii1989}
Lii,~J.~H.; Allinger,~N.~L. Molecular Mechanics. The MM3 Force Field for
  Hydrocarbons. 3. The van der Waals Potentials and Crystal Data for Aliphatic
  and Aromatic Hydrocarbons. \emph{J. Am. Chem. Soc.} \textbf{1989},
  \emph{111}, 8576--8582\relax
\mciteBstWouldAddEndPuncttrue
\mciteSetBstMidEndSepPunct{\mcitedefaultmidpunct}
{\mcitedefaultendpunct}{\mcitedefaultseppunct}\relax
\EndOfBibitem
\bibitem[Allinger \latin{et~al.}(1994)Allinger, Zhou, and
  Bergsma]{allinger1994}
Allinger,~N.~L.; Zhou,~X.; Bergsma,~J. Molecular Mechanics Parameters. \emph{J.
  Mol. Struc.-THEOCHEM} \textbf{1994}, \emph{312}, 69--83\relax
\mciteBstWouldAddEndPuncttrue
\mciteSetBstMidEndSepPunct{\mcitedefaultmidpunct}
{\mcitedefaultendpunct}{\mcitedefaultseppunct}\relax
\EndOfBibitem
\bibitem[Sun(1998)]{tailcorr}
Sun,~H. COMPASS: An ab Initio Force-Field Optimized for Condensed-Phase
  Applications - Overview with Details on Alkane and Benzene Compounds.
  \emph{J. Phys. Chem. B} \textbf{1998}, \emph{102}, 7338--7364\relax
\mciteBstWouldAddEndPuncttrue
\mciteSetBstMidEndSepPunct{\mcitedefaultmidpunct}
{\mcitedefaultendpunct}{\mcitedefaultseppunct}\relax
\EndOfBibitem
\bibitem[Dubbeldam \latin{et~al.}(2016)Dubbeldam, Calero, Ellis, and
  Snurr]{raspa}
Dubbeldam,~D.; Calero,~S.; Ellis,~D.~E.; Snurr,~R.~Q. RASPA: molecular
  simulation software for adsorption and diffusion in flexible nanoporous
  materials. \emph{Mol. Simul.} \textbf{2016}, \emph{42}, 81--101\relax
\mciteBstWouldAddEndPuncttrue
\mciteSetBstMidEndSepPunct{\mcitedefaultmidpunct}
{\mcitedefaultendpunct}{\mcitedefaultseppunct}\relax
\EndOfBibitem
\bibitem[Plimpton(1995)]{lammps}
Plimpton,~S. Fast Parallel Algorithms for Short-Range Molecular Dynamics.
  \emph{J. Comput. Phys.} \textbf{1995}, \emph{117}, 1--19\relax
\mciteBstWouldAddEndPuncttrue
\mciteSetBstMidEndSepPunct{\mcitedefaultmidpunct}
{\mcitedefaultendpunct}{\mcitedefaultseppunct}\relax
\EndOfBibitem
\bibitem[Nos\'{e}(1984)]{nose1}
Nos\'{e},~S. {A molecular dynamics method for simulations in the canonical
  ensemble}. \emph{Mol. Phys.} \textbf{1984}, \emph{52}, 255--268\relax
\mciteBstWouldAddEndPuncttrue
\mciteSetBstMidEndSepPunct{\mcitedefaultmidpunct}
{\mcitedefaultendpunct}{\mcitedefaultseppunct}\relax
\EndOfBibitem
\bibitem[Hoover(1985)]{nose2}
Hoover,~W.~G. {Canonical Dynamics: Equilibrium Phase-Space Distributions}.
  \emph{Phys. Rev. A} \textbf{1985}, \emph{31}, 1695--1697\relax
\mciteBstWouldAddEndPuncttrue
\mciteSetBstMidEndSepPunct{\mcitedefaultmidpunct}
{\mcitedefaultendpunct}{\mcitedefaultseppunct}\relax
\EndOfBibitem
\bibitem[Martyna \latin{et~al.}(1992)Martyna, Klein, and Tuckerman]{nose3}
Martyna,~G.~J.; Klein,~M.~L.; Tuckerman,~M. {Nose-Hoover Chains - the Canonical
  Ensemble Via Continuous Dynamics}. \emph{J. Chem. Phys} \textbf{1992},
  \emph{97}, 2635--2643\relax
\mciteBstWouldAddEndPuncttrue
\mciteSetBstMidEndSepPunct{\mcitedefaultmidpunct}
{\mcitedefaultendpunct}{\mcitedefaultseppunct}\relax
\EndOfBibitem
\bibitem[Martyna \latin{et~al.}(1994)Martyna, Tobias, and Klein]{mtk1}
Martyna,~G.~J.; Tobias,~D.~J.; Klein,~M.~L. {Constant-Pressure
  Molecular-Dynamics Algorithms}. \emph{J. Chem. Phys.} \textbf{1994},
  \emph{101}, 4177--4189\relax
\mciteBstWouldAddEndPuncttrue
\mciteSetBstMidEndSepPunct{\mcitedefaultmidpunct}
{\mcitedefaultendpunct}{\mcitedefaultseppunct}\relax
\EndOfBibitem
\bibitem[Martyna \latin{et~al.}(1996)Martyna, Tuckerman, Tobias, and
  Klein]{mtk2}
Martyna,~G.~J.; Tuckerman,~M.~E.; Tobias,~D.~J.; Klein,~M.~L. Explicit
  reversible integrators for extended systems dynamics. \emph{Mol. Phys.}
  \textbf{1996}, \emph{87}, 1117--1157\relax
\mciteBstWouldAddEndPuncttrue
\mciteSetBstMidEndSepPunct{\mcitedefaultmidpunct}
{\mcitedefaultendpunct}{\mcitedefaultseppunct}\relax
\EndOfBibitem
\bibitem[Ceriotti \latin{et~al.}(2010)Ceriotti, Parrinello, Markland, and
  Manolopoulos]{ceri+10jcp}
Ceriotti,~M.; Parrinello,~M.; Markland,~T.~E.; Manolopoulos,~D.~E. {Efficient
  stochastic thermostatting of path integral molecular dynamics.} \emph{J.
  Chem. Phys.} \textbf{2010}, \emph{133}, 124104\relax
\mciteBstWouldAddEndPuncttrue
\mciteSetBstMidEndSepPunct{\mcitedefaultmidpunct}
{\mcitedefaultendpunct}{\mcitedefaultseppunct}\relax
\EndOfBibitem
\bibitem[Bussi and Parrinello(2007)Bussi, and Parrinello]{buss-parr07pre}
Bussi,~G.; Parrinello,~M. {Accurate sampling using Langevin dynamics}.
  \emph{Phys. Rev. E} \textbf{2007}, \emph{75}, 56707\relax
\mciteBstWouldAddEndPuncttrue
\mciteSetBstMidEndSepPunct{\mcitedefaultmidpunct}
{\mcitedefaultendpunct}{\mcitedefaultseppunct}\relax
\EndOfBibitem
\bibitem[Bussi \latin{et~al.}(2009)Bussi, Zykova-Timan, and Parrinello]{bzp}
Bussi,~G.; Zykova-Timan,~T.; Parrinello,~M. Isothermal-isobaric molecular
  dynamics using stochastic velocity rescaling. \emph{J. Chem. Phys.}
  \textbf{2009}, \emph{130}, 074101\relax
\mciteBstWouldAddEndPuncttrue
\mciteSetBstMidEndSepPunct{\mcitedefaultmidpunct}
{\mcitedefaultendpunct}{\mcitedefaultseppunct}\relax
\EndOfBibitem
\bibitem[Lamaire \latin{et~al.}(submitted)Lamaire, Wieme, Rogge, Waroquier, and
  Van~Speybroeck]{lamaire2018}
Lamaire,~A.; Wieme,~J.; Rogge,~S. M.~J.; Waroquier,~M.; Van~Speybroeck,~V. On
  the importance of anharmonicities and nuclear quantum effects in modelling
  the structural properties and thermal expansion in {MOF}-5.
  \textbf{submitted}, \relax
\mciteBstWouldAddEndPunctfalse
\mciteSetBstMidEndSepPunct{\mcitedefaultmidpunct}
{}{\mcitedefaultseppunct}\relax
\EndOfBibitem
\bibitem[Balog \latin{et~al.}(2000)Balog, Hughes, and Martyna]{Balog2000}
Balog,~E.; Hughes,~A.~L.; Martyna,~G.~J. Constant pressure path integral
  molecular dynamics studies of quantum effects in the liquid state properties
  of n-alkanes. \emph{J. Chem. Phys.} \textbf{2000}, \emph{112}, 870--880\relax
\mciteBstWouldAddEndPuncttrue
\mciteSetBstMidEndSepPunct{\mcitedefaultmidpunct}
{\mcitedefaultendpunct}{\mcitedefaultseppunct}\relax
\EndOfBibitem
\bibitem[Pereyaslavets \latin{et~al.}(2018)Pereyaslavets, Kurnikov, Kamath,
  Butin, Illarionov, Leontyev, Olevanov, Levitt, Kornberg, and
  Fain]{Pereyaslavets2018}
Pereyaslavets,~L.; Kurnikov,~I.; Kamath,~G.; Butin,~O.; Illarionov,~A.;
  Leontyev,~I.; Olevanov,~M.; Levitt,~M.; Kornberg,~R.~D.; Fain,~B. On the
  importance of accounting for nuclear quantum effects in ab initio calibrated
  force fields in biological simulations. \emph{Proc. Natl. Acad. Sci. U.S.A.}
  \textbf{2018}, \emph{115}, 8878--8882\relax
\mciteBstWouldAddEndPuncttrue
\mciteSetBstMidEndSepPunct{\mcitedefaultmidpunct}
{\mcitedefaultendpunct}{\mcitedefaultseppunct}\relax
\EndOfBibitem
\bibitem[Veit \latin{et~al.}(2018)Veit, Jain, Bonakala, Rudra, Hohl, and
  Cs\'{a}nyi]{veit2018}
Veit,~M.; Jain,~S.~K.; Bonakala,~S.; Rudra,~I.; Hohl,~D.; Cs\'{a}nyi,~G.
  {Equation of state of fluid methane from first principles with machine
  learning potentials}. \emph{arXiv preprint arXiv:1810.10475v1} \textbf{2018},
  \relax
\mciteBstWouldAddEndPunctfalse
\mciteSetBstMidEndSepPunct{\mcitedefaultmidpunct}
{}{\mcitedefaultseppunct}\relax
\EndOfBibitem
\bibitem[Ravikovitch and Neimark(2006)Ravikovitch, and
  Neimark]{ravikovitch2006}
Ravikovitch,~P.~I.; Neimark,~A.~V. Density Functional Theory Model of
  Adsorption Deformation. \emph{Langmuir} \textbf{2006}, \emph{22},
  10864--10868\relax
\mciteBstWouldAddEndPuncttrue
\mciteSetBstMidEndSepPunct{\mcitedefaultmidpunct}
{\mcitedefaultendpunct}{\mcitedefaultseppunct}\relax
\EndOfBibitem
\bibitem[Joo \latin{et~al.}(2013)Joo, Kim, and Han]{joo2013}
Joo,~J.; Kim,~H.; Han,~S.~S. Volume shrinkage of a metal-organic framework host
  induced by the dispersive attraction of guest gas molecules. \emph{Phys.
  Chem. Chem. Phys.} \textbf{2013}, \emph{15}, 18822--18826\relax
\mciteBstWouldAddEndPuncttrue
\mciteSetBstMidEndSepPunct{\mcitedefaultmidpunct}
{\mcitedefaultendpunct}{\mcitedefaultseppunct}\relax
\EndOfBibitem
\bibitem[Zhou \latin{et~al.}(2008)Zhou, Wu, Yildirim, Simpson, and
  Hight~Walker]{zhou2008}
Zhou,~W.; Wu,~H.; Yildirim,~T.; Simpson,~J.~R.; Hight~Walker,~A.~R. Origin of
  the exceptional negative thermal expansion in metal-organic framework-5
  {Z}n$_4${O}(1,4-benzenedicarboxylate)$_3$. \emph{Phys. Rev. B} \textbf{2008},
  \emph{78}, 054114\relax
\mciteBstWouldAddEndPuncttrue
\mciteSetBstMidEndSepPunct{\mcitedefaultmidpunct}
{\mcitedefaultendpunct}{\mcitedefaultseppunct}\relax
\EndOfBibitem
\bibitem[Lock \latin{et~al.}(2010)Lock, Wu, Christensen, Cameron, Peterson,
  Bridgeman, Kepert, and Iversen]{lock2010}
Lock,~N.; Wu,~Y.; Christensen,~M.; Cameron,~L.~J.; Peterson,~V.~K.;
  Bridgeman,~A.~J.; Kepert,~C.~J.; Iversen,~B.~B. Elucidating Negative Thermal
  Expansion in MOF-5. \emph{J. Phys. Chem. C} \textbf{2010}, \emph{114},
  16181--16186\relax
\mciteBstWouldAddEndPuncttrue
\mciteSetBstMidEndSepPunct{\mcitedefaultmidpunct}
{\mcitedefaultendpunct}{\mcitedefaultseppunct}\relax
\EndOfBibitem
\bibitem[Lock \latin{et~al.}(2013)Lock, Christensen, Wu, Peterson, Thomsen,
  Plitz, Ramirez-Cuesta, McIntyre, Nor\'{e}n, Kutteh, Kepert, Kearley, and
  Iversen]{lock2013}
Lock,~N.; Christensen,~M.; Wu,~Y.; Peterson,~V.~K.; Thomsen,~M.~K.;
  Plitz,~R.~O.; Ramirez-Cuesta,~A.~J.; McIntyre,~G.~J.; Nor\'{e}n,~K.;
  Kutteh,~R.; Kepert,~C.~J.; Kearley,~G.~J.; Iversen,~B.~B. Scrutinizing
  negative thermal expansion in MOF-5 by scattering techniques and ab initio
  calculations. \emph{Dalton Trans.} \textbf{2013}, \emph{42}, 1996--2007\relax
\mciteBstWouldAddEndPuncttrue
\mciteSetBstMidEndSepPunct{\mcitedefaultmidpunct}
{\mcitedefaultendpunct}{\mcitedefaultseppunct}\relax
\EndOfBibitem
\bibitem[not()]{note-1}
A 10 $\%$ increase in volume due to NQEs, assuming the molecules to be
  spherical is associated with a $(1.10)^{\frac{1}{3}}$ times increase in the
  effective radius. The classical "radius" is close to 4 \AA. And thus the
  increase in the inter-molecular distance due to NQEs coming from the
  isotropic expansion of the gas is $(1.10)^{\frac{1}{3}} * 4 - 4 \approx 0.13$
  \AA.\relax
\mciteBstWouldAddEndPunctfalse
\mciteSetBstMidEndSepPunct{\mcitedefaultmidpunct}
{}{\mcitedefaultseppunct}\relax
\EndOfBibitem
\bibitem[Ming \latin{et~al.}(2014)Ming, Purewal, Liu, Sudik, Xu, Yang,
  Veenstra, Rhodes, Soltis, Warner, Gaab, M\"{u}ller, and Siegel]{ming2014}
Ming,~Y.; Purewal,~J.; Liu,~D.; Sudik,~A.; Xu,~C.; Yang,~J.; Veenstra,~M.;
  Rhodes,~K.; Soltis,~R.; Warner,~J.; Gaab,~M.; M\"{u}ller,~U.; Siegel,~D.~J.
  Thermophysical properties of MOF-5 powders. \emph{Microporous Mesoporous
  Mater.} \textbf{2014}, \emph{185}, 235--244\relax
\mciteBstWouldAddEndPuncttrue
\mciteSetBstMidEndSepPunct{\mcitedefaultmidpunct}
{\mcitedefaultendpunct}{\mcitedefaultseppunct}\relax
\EndOfBibitem
\bibitem[Liu \latin{et~al.}(2012)Liu, Purewal, Yang, Sudik, Maurer, M\"{u}ller,
  Ni, and Siegel]{liu2012}
Liu,~D.; Purewal,~J.~J.; Yang,~J.; Sudik,~A.; Maurer,~S.; M\"{u}ller,~U.;
  Ni,~J.; Siegel,~D.~J. MOF-5 composites exhibiting improved thermal
  conductivity. \emph{Int. J. Hydrogen Ener.} \textbf{2012}, \emph{37},
  6109--6117\relax
\mciteBstWouldAddEndPuncttrue
\mciteSetBstMidEndSepPunct{\mcitedefaultmidpunct}
{\mcitedefaultendpunct}{\mcitedefaultseppunct}\relax
\EndOfBibitem
\bibitem[Addicoat \latin{et~al.}(2014)Addicoat, Vankova, Akter, and
  Heine]{addicoat2014}
Addicoat,~M.~A.; Vankova,~N.; Akter,~I.~F.; Heine,~T. Extension of the
  Universal Force Field to Metal-Organic Frameworks. \emph{J. Chem. Theory
  Comput.} \textbf{2014}, \emph{10}, 880--891\relax
\mciteBstWouldAddEndPuncttrue
\mciteSetBstMidEndSepPunct{\mcitedefaultmidpunct}
{\mcitedefaultendpunct}{\mcitedefaultseppunct}\relax
\EndOfBibitem
\bibitem[Wilmer \latin{et~al.}(2012)Wilmer, Leaf, Lee, Farha, Hauser, Hupp, and
  Snurr]{wilmer2012}
Wilmer,~C.~E.; Leaf,~M.; Lee,~C.~Y.; Farha,~O.~K.; Hauser,~B.~G.; Hupp,~J.~T.;
  Snurr,~R.~Q. Large-scale screening of hypothetical metal-organic frameworks.
  \emph{Nat. Chem.} \textbf{2012}, \emph{4}, 83--89\relax
\mciteBstWouldAddEndPuncttrue
\mciteSetBstMidEndSepPunct{\mcitedefaultmidpunct}
{\mcitedefaultendpunct}{\mcitedefaultseppunct}\relax
\EndOfBibitem
\bibitem[Moghadam \latin{et~al.}(2018)Moghadam, Islamoglu, Goswami, Exley,
  Fantham, Kaminski, Snurr, Farha, and Fairen-Jimenez]{moghadam2018}
Moghadam,~P.~Z.; Islamoglu,~T.; Goswami,~S.; Exley,~J.; Fantham,~M.;
  Kaminski,~C.~F.; Snurr,~R.~Q.; Farha,~O.~K.; Fairen-Jimenez,~D.
  Computer-aided discovery of a metal-organic framework with superior oxygen
  uptake. \emph{Nat. Commun.} \textbf{2018}, \emph{9}, 1378\relax
\mciteBstWouldAddEndPuncttrue
\mciteSetBstMidEndSepPunct{\mcitedefaultmidpunct}
{\mcitedefaultendpunct}{\mcitedefaultseppunct}\relax
\EndOfBibitem
\bibitem[Tabor \latin{et~al.}(2018)Tabor, Roch, Saikin, Kreisbeck, Sheberla,
  Montoya, Dwaraknath, Aykol, Ortiz, Tribukait, Amador-Bedolla, Brabec,
  Maruyama, Persson, and A.]{tabor2018}
Tabor,~D.~P.; Roch,~L.~M.; Saikin,~S.~K.; Kreisbeck,~C.; Sheberla,~D.;
  Montoya,~J.~H.; Dwaraknath,~S.; Aykol,~M.; Ortiz,~C.; Tribukait,~H.;
  Amador-Bedolla,~C.; Brabec,~C.~J.; Maruyama,~B.; Persson,~K.~A.; A.,~A.-G.
  Accelerating the discovery of materials for clean energy in the era of smart
  automation. \emph{Nat. Rev. Mater.} \textbf{2018}, \emph{3}, 5--20\relax
\mciteBstWouldAddEndPuncttrue
\mciteSetBstMidEndSepPunct{\mcitedefaultmidpunct}
{\mcitedefaultendpunct}{\mcitedefaultseppunct}\relax
\EndOfBibitem
\bibitem[Tsivion and Head-Gordon(2017)Tsivion, and Head-Gordon]{tsivion2017}
Tsivion,~E.; Head-Gordon,~M. Methane Storage: Molecular Mechanisms Underlying
  Room-Temperature Adsorption in {Z}n$_4${O}({BDC})$_3$ ({MOF}-5). \emph{J.
  Phys. Chem. C} \textbf{2017}, \emph{121}, 12091--12100\relax
\mciteBstWouldAddEndPuncttrue
\mciteSetBstMidEndSepPunct{\mcitedefaultmidpunct}
{\mcitedefaultendpunct}{\mcitedefaultseppunct}\relax
\EndOfBibitem
\bibitem[Wu \latin{et~al.}(2009)Wu, Zhou, and Yildirim]{wu2009}
Wu,~H.; Zhou,~W.; Yildirim,~T. Methane Sorption in Nanoporous Metal-Organic
  Frameworks and First-Order Phase Transition of Confined Methane. \emph{J.
  Phys. Chem. C} \textbf{2009}, \emph{113}, 3029--3035\relax
\mciteBstWouldAddEndPuncttrue
\mciteSetBstMidEndSepPunct{\mcitedefaultmidpunct}
{\mcitedefaultendpunct}{\mcitedefaultseppunct}\relax
\EndOfBibitem
\bibitem[Kuchta \latin{et~al.}(2017)Kuchta, Dundar, Formalik, Llewellyn, and
  Firley]{kuchta2017}
Kuchta,~B.; Dundar,~E.; Formalik,~F.; Llewellyn,~P.~L.; Firley,~L.
  Adsorption-Induced Structural Phase Transformation in Nanopores. \emph{Angew.
  Chem. Int. Ed.} \textbf{2017}, \emph{56}, 16243--16246\relax
\mciteBstWouldAddEndPuncttrue
\mciteSetBstMidEndSepPunct{\mcitedefaultmidpunct}
{\mcitedefaultendpunct}{\mcitedefaultseppunct}\relax
\EndOfBibitem
\bibitem[Ceriotti \latin{et~al.}(2009)Ceriotti, Bussi, and
  Parrinello]{ceri+09prl}
Ceriotti,~M.; Bussi,~G.; Parrinello,~M. {Langevin Equation with Colored Noise
  for Constant-Temperature Molecular Dynamics Simulations}. \emph{Phys. Rev.
  Lett.} \textbf{2009}, \emph{102}, 020601\relax
\mciteBstWouldAddEndPuncttrue
\mciteSetBstMidEndSepPunct{\mcitedefaultmidpunct}
{\mcitedefaultendpunct}{\mcitedefaultseppunct}\relax
\EndOfBibitem
\bibitem[Habershon \latin{et~al.}(2009)Habershon, Markland, and
  Manolopoulos]{Habershon2009}
Habershon,~S.; Markland,~T.~E.; Manolopoulos,~D.~E. Competing quantum effects
  in the dynamics of a flexible water model. \emph{The Journal of Chemical
  Physics} \textbf{2009}, \emph{131}, 024501\relax
\mciteBstWouldAddEndPuncttrue
\mciteSetBstMidEndSepPunct{\mcitedefaultmidpunct}
{\mcitedefaultendpunct}{\mcitedefaultseppunct}\relax
\EndOfBibitem
\end{mcitethebibliography}
\end{document}